\documentclass{article}
\usepackage{amsmath}
\usepackage{amsfonts}
\usepackage{amssymb}
\usepackage{graphicx}
\usepackage{overpic}
\usepackage{lastpage}
\usepackage[margin=1.0in]{geometry}
\usepackage{fancyhdr, lastpage}
\usepackage[flushleft]{threeparttable}
\usepackage{float}
\usepackage{epsfig}
\usepackage{tabularx}
\usepackage[font=small,labelfont=bf]{caption}
\usepackage{xcolor}
\usepackage{titling}
\usepackage{cite}
\usepackage{soul}
\usepackage{cancel}
\usepackage{subcaption}
\usepackage{bm}
\usepackage{svg}
\usepackage[colorlinks = true,
            linkcolor = blue,
            urlcolor  = blue,
            citecolor = blue,
            anchorcolor = blue]{hyperref}
\begin{document}
%\pagenumbering{gobble}
%\pagenumber
%\Large
% \begin{center}
\title{\vspace{-1.15cm} $T_i/T_e$ Dependence of Core Turbulence and Transport in DIII-D QH-Mode Plasmas \vspace{-1.5em}}
\date{}
\maketitle
\vspace{-1.5cm}
\begin{center}
\fontsize{10}{15}\selectfont Abhishek Tiwari$^{1,*}$, Kshitish Barada$^2$, Jaya Kumar Alageshan$^{1}$, Santanu Banerjee$^3$, Tanmay Macwan$^2$, Terry L. Rhodes$^2$, Sarveshwar Sharma$^{4,5}$, Zhihong Lin$^{6}$ , and Animesh Kuley$^{1,*}$  \\
\vspace{0.25cm}
\small \textit{$^{1}$Department of Physics, Indian Institute of Science, Bangalore 560012, India} \\
\small \textit{$^{2}$Physics and Astronomy Department, University of California, Los Angeles, California 90995, USA} \\
\small \textit{$^{3}$Princeton Plasma Physics Laboratory, Princeton, New Jersey 08543, USA} \\
\small \textit{$^4$Institute for Plasma Research, Bhat, Gandhinagar 382428, India} \\
\small \textit{$^5$Homi Bhabha National Institute, Anushaktinagar, Mumbai, Maharashtra 400094, India}\\
\small \textit{$^6$Department of Physics and Astronomy, University of California Irvine, CA 92697, USA}\\
\textit{$^{*}$Email: abhishektiwa@iisc.ac.in, akuley@iisc.ac.in}\\
\end{center}

\begin{abstract}

\noindent This study investigates the effect of the ion-to-electron temperature ratio ($T_i/T_e$) on microturbulence driven transport in Quiescent H-mode (QH-mode) plasmas in the DIII-D tokamak. Utilizing the Gyrokinetic Toroidal Code (GTC) and the QH-mode equilibrium, we perform linear and nonlinear simulations to analyze transport properties and instability dynamics under variations of $T_i$ and $T_e$. Our results demonstrate that decreasing $T_i/T_e$ leads to a relative destabilization of trapped electron modes (TEM) over ion temperature gradient (ITG) modes, with the transition between these regimes dictated by $T_i/T_e$. When the electron temperature is increased at fixed ion temperature, we observe an increase in transport saturation levels. In contrast, decreasing the ion temperature at fixed electron temperature results in more modest transport enhancement. The radial correlation length, which characterizes eddy size, increases with rising $T_e$ and decreases with falling $T_i$, consistent with the observed trends in turbulent transport. Additionally, we examine the impact of impurity addition on turbulence and growth rates, finding that impurity presence does not significantly alter transport quantities compared to the impurity-free case. Finally, investigating helium as an alternative main ion species, we find that helium plasmas exhibit higher linear growth rates but result in lower transport saturation levels than deuterium plasmas, suggesting potential confinement benefits. These findings provide quantitative insights into the temperature ratio dependence in QH-mode plasmas and highlight the role of temperature profiles and zonal flows in influencing plasma confinement.
\end{abstract}

\textit{Keywords:} Simulations, Gyrokinetic, Microturbulence, Tokamak, QH-Mode Plasma
%\newpage

\section{\fontsize{12}{15}\label{sec:intro}Introduction}
High-confinement (H-mode) plasmas are an important operational scenario in future fusion devices\cite{groebner1993emerging}, characterized by steep edge temperature and density gradients. However, these steep gradients lead to the occurrence of Edge Localized modes (ELMs)\cite{zohm1996edge,wilson2006magneto}. ELMs are the peeling-ballooning MHD instability that degrades H-mode plasma performance in tokamaks\cite{viezzer2018access}. ELMs pose a significant challenge, as they expel particles and energy from the plasma, leading to transient degradation of the transport barrier\cite{viezzer2018access}. The heat and particles expelled by ELMs can also damage plasma-facing components, which can reduce their lifetime\cite{blanchard2018effect}. Therefore, suppressing ELMs is essential. Various strategies are used to eliminate ELMs, such as resonant magnetic perturbation (RMP) ELM suppression, pellet pacing to increase ELM-frequency,  naturally ELM-free operation like I-mode, and QH mode\cite{grierson2015impurity,maingi2014enhanced,lang2013elm}. More recently, alternative plasma shaping, notably negative triangularity, has been shown to reduce or eliminate large type-I ELMs in some conditions \cite{austin2019achievement,thome2024overview}. Other ELM-mitigation or ELM-free regimes - for example, the enhanced-D$\alpha$ (EDA) regime\cite{greenwald1999characterization, mossessian2003edge, paz2021plasma,kalis2023experimental} and quiescent-confinement variants such as Quasi-Continuous Exhaust (QCE)\cite{kirk2011comparison,labit2019dependence,saibene2005characterization,ozeki1990plasma,xu2019promising,kamada2000disappearance} have also been demonstrated on multiple devices. Techniques currently used to suppress ELMs in DIII-D include RMP ELM suppression\cite{evans2006edge,grierson2015impurity} and Quiescent H-mode (QH-mode) plasma operation\cite{burrell2002quiescent}. RMP ELM suppression relies on externally applied magnetic fields to enhance edge particle transport and mitigate ELMs. On the other hand, QH-mode plasma is a steady and naturally ELM-free regime\cite{viezzer2018access}. The QH-mode plasma was discovered in DIII-D tokamak in 1999\cite{burrell2002quiescent} and later observed in other tokamaks like ASDEX\cite{suttrop2004study}, JET-C\cite{suttrop2005studies}, and JT-60U\cite{sakamoto2004impact}. The natural stability of QH-mode plasma against ELMs arises due to Edge Harmonic Oscillations (EHO)\cite{burrell2002quiescent}, which results in an increased edge particle transport and has been shown to aid in the transport of both main ions and impurity ions\cite{grierson2015impurity}. This additional particle transport driven by the EHO allows the plasma edge to reach a transport equilibrium at edge pressures and current densities just below the ELM stability boundary\cite{garofalo2011advances,greenfield2001quiescent,snyder2007stability}.

The ion-to-electron temperature ratio ($T_i/T_e$) critically influences plasma confinement and impacts anomalous transport, as demonstrated in previous experiments\cite{fontana2019effects,barada2021new,banerjee2024decoupling,banerjee2021evolution}. In future fusion devices like ITER, long energy confinement times, compared to electron-ion thermal equilibration times, are expected to bring ion and electron temperatures closer. In this context, understanding $T_i/T_e$ behavior becomes particularly relevant for operational regimes like the QH-mode, which are naturally ELM-free and thus offer advantages for reducing ELM-induced damage in reactor-scale devices. However, their successful projection to burning plasma conditions requires a detailed understanding of transport properties. The $T_i/T_e$ ratio is a key parameter influencing turbulence characteristics and encapsulates the electron-ion heat exchange, directly affecting transport dynamics. Experimentally, $T_i/T_e$ can be varied by leveraging preferential heating methods—Electron Cyclotron Heating (ECH) to raise electron temperature, and Neutral Beam Injection (NBI) or Ion Cyclotron Resonance Heating (ICRH) for ion heating—under different electron-ion collisionalities. These heating techniques--NBI, ICRH, and ECH--remain essential tools for modifying the $T_i/T_e$ ratio, with NBI and ICRH primarily used for ion heating, current drive, and momentum injection, while ECH provides precise control over electron temperature and current profiles\cite{hopf2021neutral,krasilnikov2009fundamental}. We note, however, that while ICRH effectively heats ions, the use of ICRH alone has not been shown to supply the edge torque normally required for accessing QH-mode on devices such as DIII-D, where reliable QH-mode operation typically depends on strong counter-NBI momentum input\cite{burrell2002quiescent}.

Future fusion devices like ITER will employ a broader set of heaters (NBI, ECH and ICRH) in its baseline configuration\cite{ITER2024}. Since QH-mode is operated in a low-collisionality regime\cite{garofalo2011advances}, ion-to-electron temperature equilibration is hindered. To successfully project QH-mode as an ELM-free plasma regime in future fusion devices, it is essential to understand the effects of different heating mechanisms on transport and stability. We begin by investigating transport in QH-mode plasmas of DIII-D by systematically varying the $T_i/T_e$ ratio, which effectively simulates varying the heating mix in experiments.

Several studies have investigated the influence of the $T_i/T_e$ ratio across various tokamaks, including DIII-D\cite{mckee2014turbulence,ernst2016role,mordijck2015particle,petty1999dependence,stallard1999electron,houshmandyar2022explaining}, KSTAR\cite{hong2015control}, JET\cite{jet1999alpha}, C-Mod\cite{ernst2014controlling}, JT-60U\cite{yoshida2013temporal}, and AUG\cite{ryter2019heat,sommer2015transport}. Studies have shown that the $T_i/T_e$ ratio affects the ITG threshold by influencing its stability, leading to either destabilization or suppression depending on the specific plasma conditions \cite{gao2001effects,petty1999dependence}. Previous studies~\cite{mckee2014turbulence,yoshida2017magnetic,sommer2015transport} have observed confinement degradation as the $T_i/T_e$ ratio is decreased via ECH due to increased transport quantities in all channels. Moreover, \cite{he2024ti} presents a study on the dependence of confinement on the $T_i/T_e$ ratio in low $q_{95}$ plasmas, demonstrating that a higher $T_i/T_e$ enhances confinement by suppressing ITG-mode turbulence. Most of these studies focus on H-mode plasmas. Since future fusion devices are expected to operate in ELM-free regimes, investigating the impact of the $T_i/T_e$  ratio on transport in QH-mode plasmas is essential for optimizing confinement and improving tokamak performance.

While $T_i/T_e$ influences transport properties, impurities also play a significant role in plasma performance and must be controlled to optimize fusion efficiency. Impurities are unavoidable in fusion reactors due to various processes, such as the production of helium ash during plasma burning, and the sputtering of wall materials due to plasma-wall interactions\cite{bohdansky1981plasma}. The presence of impurities leads to fuel dilution, which reduces fusion power\cite{coppi1966drift,jensen1977calculations,isler1984impurities,singh2024gyrokinetic}. A higher impurity concentration could lead to radiation losses via bremsstrahlung and recombination radiation\cite{hinnov1978effects,putterich2008modelling}. However, a small amount of impurity can be beneficial for confinement. For instance, impurities have been observed to stabilize the ITG mode due to ion dilution in JET\cite{bonanomi2018effects}, WEST\cite{yang2020core}, and improvement in confinement in ASDEX\cite{tardini2012core,fable2021high}, W7-X\cite{lunsford2021characterization}, DIII-D\cite{mckee2000impurity,murakami2001physics, PhysRevLett.84.1922}, and LHD\cite{nespoli2022observation}. In contrast, a higher concentration of impurities can be detrimental to the confinement\cite{angioni2021impurity,dong1995studies}. QH-mode plasma has been observed with modest carbon impurity in DIII-D\cite{ma2022characterizing} (where the walls are made of graphite).

Beyond impurities, the choice of main ions—such as helium—and the selection of hydrogen isotopes (H/D/T) can also influence turbulence and transport behavior in fusion plasmas \cite{maggi2017isotope}. In preparation for ITER’s early operational phases, helium plasmas are of particular interest due to their role in non-nuclear startup scenarios and their naturally lower L–H transition power threshold\cite{ITER2024}. While helium offers practical advantages, challenges such as ELM sustainment and core–edge integration remain. As QH-mode plasmas are naturally ELM-free, investigating core transport in helium QH-mode regimes is both timely and important.

This paper is structured as follows: Section~\ref{sec:simmodel} introduces the simulation model and describes the gyrokinetic Vlasov and Poisson equations used in the GTC code to model microturbulence-driven transport. Section~\ref{sec: experimentalsetup} outlines the experimental setup, including the temperature and density profiles. Sections~\ref{sec: simDion} and~\ref{sec: simele} investigate the impact of the $T_i/T_e$ ratio on transport by varying the ion and electron temperatures independently using gyrokinetic simulations, while holding the other fixed. Section~\ref{sec:comparetwostudies} compares the findings from these two cases, and Section~\ref{sec:impstudy} examines the influence of impurities in the presence of zonal flows. We also compare the GTC simulation results with TRANSP, an interpretive code that infers experimental ion, electron and particle transport coefficients from measured profiles and power balance; this provides an experimental benchmark for the first-principles, gyrokinetic predictions of GTC. Finally, Section~\ref{sec:heplasma} analyzes transport behavior in QH-mode plasmas using helium as the main ion species.

\section{\fontsize{12}{15}\selectfont \label{sec:simmodel} Simulation Model}
\noindent We use the global gyrokinetic code GTC\cite{lin1998turbulent} to perform collisionless 
gyrokinetic simulations of microturbulence transport. GTC has been extensively applied to study microturbulent transport of tokamaks \cite{singh2024gyrokinetic,Singh_2023,xiao2015gyrokinetic,Tajinder2025} and stellarators \cite{singh2022global,singh2023global,tiwari2025,wang2020global}, Alfv\'en eigenmodes and MHD instabilities\cite{PhysRevLett.111.145003,liu2022regulation,brochard2024saturation}, energetic particles transport\cite{PhysRevLett.101.095001}, and radio frequency waves\cite{Bao_2014,Kuley15}. GTC takes advantage of the anisotropic nature of microturbulence, $k_{||} \ll k_{\perp}$, to reduce the computational cost by using a field-aligned mesh to represent fluctuation quantities, which helps maintain the numerical and computational efficiency of the simulations.

The dynamics of thermal and impurity ions are governed by the 5-D gyrokinetic Vlasov equation\cite{singh2022global,singh2023global,brizard2007foundations} in an inhomogeneous magnetic field as

\begin{eqnarray}
    \frac{d}{dt}f_{\alpha}(\Vec{X},\mu,v_{||},t) = \left[ \frac{\partial}{\partial t}+ 
    \dot{\Vec{X}} \cdot \nabla + \dot{v}_{||}\frac{\partial}{\partial v_{||}} \right]f_\alpha= 0;
    \quad\quad \dot{\Vec{X}} \; = \; v_{||}{\hat{b}} + \Vec{v}_d+ \Vec{v}_E 
    \label{eq:gyrovlas}
\end{eqnarray}
\begin{align}
\text{where,}\quad\quad\quad 
\dot{v}_{||} \; &= \; -\frac{1}{m_{\alpha}} \frac{\Vec{B}^*}{B} \cdot (\mu_{\alpha}\nabla B + Z_{\alpha}\nabla\phi) \nonumber \\
\Vec{v}_d \; &= \; \frac{v_{||}^2}{\Omega_\alpha}\; (\nabla \times {\hat{b}}) + \frac{\mu_{\alpha}}{m_\alpha\Omega_\alpha} \; ({\hat{b}} \times \nabla B) \; \nonumber\\ 
\Vec{v}_E \; &= \; \frac{c}{B} \; ({\hat{b}} \times \nabla \phi) \; \nonumber\\ \nonumber
\end{align}

\noindent where $f_{\alpha}$ is the particle distribution function, $\alpha$ labels the thermal ions, electron, and the impurity ions; $\alpha= i, e,$ and $z$ respectively. $\Vec{X}$ represents the guiding center position of the particle, $\mu_{\alpha}$ is the magnetic moment and $v_{||}$ is the velocity parallel to the magnetic field. $\Vec{v}_E$ is the $\Vec{E}\times\Vec{B}$ drift velocity, $\Vec{v}_d$ is the sum of the curvature drift and magnetic field gradient drift. $Z_{\alpha}$ and $m_{\alpha}$ label the charge and the mass of the particle. $\Vec{B}$ is the equilibrium magnetic field at the particle position and we define $\Vec{B^*}= \Vec{B} +\frac{Bv_{||}}{\Omega_{\alpha}} \nabla \times \hat{b}$ as the equilibrium magnetic field at the guiding center of the particle, and $\hat{b}= \frac{\Vec{B}}{B}$. Finally, $\phi$ is the electrostatic potential which can be decomposed into two parts as a non-zonal potential $\delta \phi$ and the zonal potential $\phi_{\text{ZF}}$ such that $\phi= \delta \phi + \phi_{\text{ZF}}$. GTC allows the zonal component $\phi_{\text{ZF}}$ to be selectively included or excluded to study its impact on microturbulence and transport.

\noindent  GTC uses low noise $\delta 
f$ method to reduce the particle noise due to Monte Carlo 
sampling of particle distribution. In this method, only the perturbed part of the particle distribution evolves with time. In this scheme, we decompose the distribution function into an
unperturbed equilibrium part and a perturbed part as $f_{\alpha}=f_{0_{\alpha}}+\delta f_{\alpha}$. Further, the propagator
in  Eq.(\ref{eq:gyrovlas}) can be separated into an equilibrium part $L_0$ and a
perturbed part, $\delta L$ so that the Eq.(\ref{eq:gyrovlas}) can be written
as $(L_0+\delta L)(f_{0_{\alpha}}+\delta f_{\alpha})=0$, where
\[L_0= \frac{\partial}{\partial t}+ (v_{||}\hat{b} + \Vec{v}_d)\cdot \nabla -\frac{1}
{m_{\alpha}} \frac{\Vec{B}^*}{B} \cdot (\mu_{\alpha}\nabla B)\frac{\partial}{\partial v_{||}},\]
\[\delta L= \Vec{v}_E\cdot\nabla -\frac{1}{m_{\alpha}}\frac{\Vec{B}^*}{B}\cdot Z_{\alpha}\nabla\phi\frac{\partial}
{\partial v_{||}}\]
The equilibrium distribution function $f_{0_{\alpha}}$ is determined by the condition $L_0 f_{0_{\alpha}}=0$. The solution of this equation is approximated to be the local 
Maxwellian
\[f_{0_{\alpha}}= \frac{n_{\alpha}}{(2\pi T_{\alpha}/m_{\alpha})^{3/2}} \exp\left( -\frac{2\mu_{\alpha} B +m_{\alpha}v_{||}^2}{2 T_{\alpha}} \right)\]

\noindent Thus, GTC only calculates the perturbed part $\delta f_{\alpha}$. GTC introduces another variable called particle weight $w_{\alpha}= \delta f_{\alpha}/f_{{\alpha}}$ which also evolves with time as\cite{singh2022global,parker1993fully}

\begin{eqnarray}
    \frac{dw_{\alpha}}{dt}= (1-w_{\alpha}) \left[ -\Vec{v_E}\cdot \frac{\nabla f_{0_{\alpha}}}{f_{0_{\alpha}}} + 
    \frac{Z_{\alpha}}{m_{\alpha} f_{0_{\alpha}}} \frac{\Vec{B^*}}{B} \cdot \nabla \phi \frac{\partial f_{0_{\alpha}}}{\partial v_{||}}
    \right]
    \label{eq:dynamicalweightevolve}
\end{eqnarray}

\noindent A kinetic treatment of electrons is necessary to describe the electron response accurately in a gyrokinetic framework. GTC implements a fluid-kinetic hybrid model to circumvent the difficulties present due to the electron parallel Courant condition and high-frequency oscillations \cite{singh2022global,Singh_2023,lin2007global}. In this model, the distribution function for the electron, $f_e$ is written as sum of three parts: unperturbed part $f_{0e}$, adiabatic part $\delta f^{(0)}_e= f_0 e^\frac{e \delta\phi^{(0)}}{T_e}$ and a non-adiabatic part $\delta g_e$ such that $f_e= f_{0e}+ f_0 e^\frac{e \delta\phi^{(0)}}{T_e} +\delta g_e $. The first term satisfies $L_0f_{0e}=0$. The electron response is adiabatic to the lowest order, whereas the non-adiabatic describes a higher-order response. The non-adiabatic parts are smaller than the adiabatic parts by a factor of $\delta$, where $\delta$ represents the fraction of magnetically trapped electrons.

The gyrokinetic Poisson equation\cite{Singh_2023,singh2024gyrokinetic,lee1987gyrokinetic} gives the electrostatic potential as

\begin{eqnarray}
    \frac{e\tau}{T_e}(\phi-\tilde{\phi})= \frac{\delta \Bar{n_i}-\delta n_e}{n_0},
    \label{eq:gyropoisson}
\end{eqnarray}

\noindent where $\delta \Bar{n_i}$, $\delta n_e$ are the ion and the electron guiding center charge density, $n_0$ is the equilibrium electron density, and $\tau= T_e/T_i$. The non-zonal component of electrostatic potential in the lowest order of electron response can be written as

\begin{eqnarray}
    \frac{(\tau+1)e \delta\phi^{(0)}}{T_e}- \frac{\tau e \delta \tilde{\phi}^{(0)}}{T_e}= \frac{\delta \Bar{n_i}-{\langle\delta \Bar{n_i}\rangle}}{n_0}
\end{eqnarray}

where we define the second gyro-averaged perturbed potential, $\delta \tilde{\phi}^{(0)}$ as 
% \cite{singh2022global}
\begin{eqnarray*}
    &\delta\tilde\phi^{(0)}(\Vec{x})= \frac{1}{2\pi}\int d^3\Vec{v} \; \int d^3\Vec{X} \; f_0(\Vec{X}) \; \delta 
    \Bar\phi^{(0)}(\Vec{X}) \; \delta(\Vec{X}+\Vec{\rho}-\Vec{x})
    \label{eq:nzgyropoiss}
\end{eqnarray*}

The first gyro-averaged perturbed potential is defined as
% \cite{singh2022global}
\begin{eqnarray*}
    \delta\Bar{\phi}^{(0)}(\Vec{X}) = \int d^3 \Vec{x}\int \frac{d\alpha}{2\pi}\delta\phi^{(0)}(\Vec{x}) \delta(\Vec{x}-\Vec{X}-\Vec{\rho})
\end{eqnarray*}

Similarly,
\begin{eqnarray*}
    \delta\Bar{n}_i(\Vec{X}) = \int d^3 \Vec{x}\int \frac{d\alpha}{2\pi}\delta f(\Vec{x}) \delta(\Vec{x}-\Vec{X}-\Vec{\rho})
\end{eqnarray*}

\noindent Here, $\Vec{x}$ and $\Vec{X}$ are the particle position and particle guiding center position coordinates, respectively, and $\Vec{\rho}$ is the gyro-radius vector. $\alpha$ is the gyro-phase. We update the particle orbits and fields iteratively in time. An RK2 solver is used to push the particles from $k$th time to $(k+1)$th time to the lowest order of non-zonal potential. 

To calculate higher-order terms of the electron distribution, GTC evolves the electron weight $w_e= \delta g_e/f_e$ satisfying
% \cite{singh2022global}

\begin{eqnarray}
    \frac{dw_{e}}{dt}= (1-w_{e}- \frac{e\delta \phi^{(0)}}{T_e}) \left[ \left. -\Vec{v}_E\cdot \frac{\nabla f_{0_e}}{f_{0_e}}  \right\vert_{\Vec{v}_\perp}
    -\frac{\partial}{\partial t}\left(\frac{e\delta \phi^{(0)}}{T_e}\right)-(\Vec{v}_d+\Vec{\delta v}_E)\cdot\nabla\left(\frac{e\phi}{T_e}\right)
    \right]
    \label{eq:dynamicalweightevolveele}
\end{eqnarray}

\noindent Here, $\delta \Vec{v}_E= (c/B^*)\hat{b}\times\nabla\delta\phi$. To solve the above equation, we first approximate the exact solution by the lowest order solution $\delta\phi^{(0)}$. All field quantities are evaluated at the $k$th time in equation~\ref{eq:dynamicalweightevolveele}, and the electron orbits are pushed to the $(k+1)$th time. The expression for the non-zonal component of electrostatic potential can now be obtained to the first order using the electron distribution function and the expansion of the electrostatic potential as

\begin{eqnarray}
    e^{e\delta\phi/T_e}= e^{e\delta\phi^{(0)}/T_e}- \frac{\delta n_e-\langle\delta n_e\rangle}{n_0}
    \label{eq:nzphiele}
\end{eqnarray}

\noindent where $\delta n_e= \int \delta g_e d^3\Vec{v}$. Depending on the trapped fraction of electrons, Equations~\ref{eq:dynamicalweightevolveele} and ~\ref{eq:nzphiele} are solved iteratively. In this way, we update all particle orbits, and the non-zonal components of field quantities are updated at $(k+1)$ time from the $k$th time. The zonal component for the electrostatic potential can be obtained by using
\begin{eqnarray}
    \frac{\tau e \left(\langle\phi\rangle -\langle\tilde\phi\rangle\right)}{T_e}= \frac{\langle\delta\bar n_i\rangle- \langle\delta n_e\rangle}{n_0}
    \label{eq:zfphiele}
\end{eqnarray}

\noindent Finally, our simulations neglect collisional effects, as the normalized collisionality is relatively low for our profile (with $\nu_* \sim 0.2$\cite{chen2016rotational}). Here, $\nu_* = \epsilon^{-3/2} \nu q R_0 / v_{\text{th}}$, where $\epsilon = r / R_0$ is the local inverse aspect ratio, $\nu$ is the physical collision frequency, and $v_{\text{th}} = \sqrt{T_{0\alpha} / m_\alpha}$ is the thermal velocity of plasma species $\alpha$. We employ a Gaussian boundary decay in our simulations for fields and energy-conserving boundary conditions for particles exiting the simulation boundary.

\begin{figure}
\includegraphics[width=0.6\textwidth]{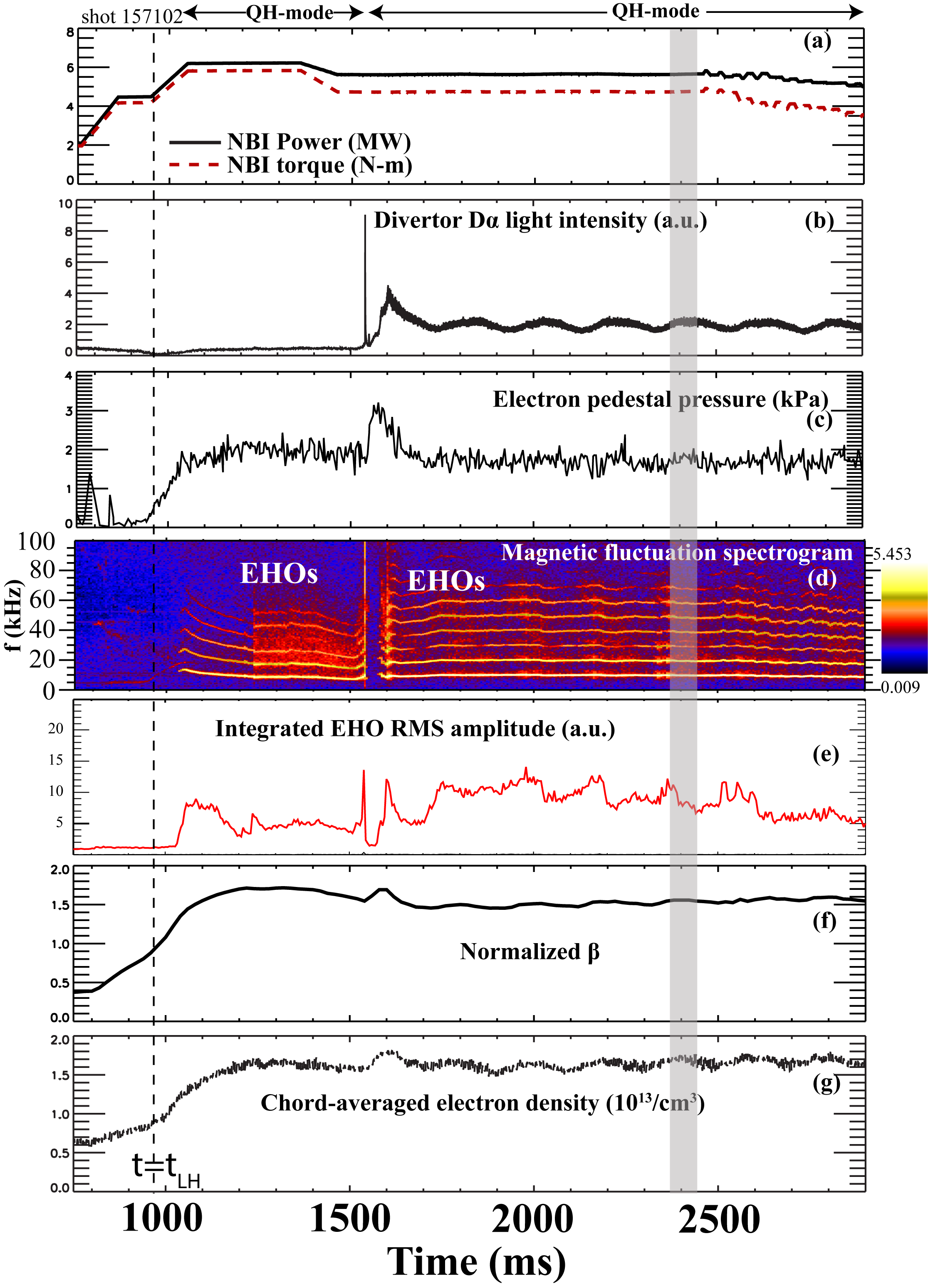}
\centering
\caption{Time evolution of plasma parameters for the discharge \#157102. We chose a small time window around $2420$ms in the above discharge marked with the grey box to simulate plasma. The QH mode operation starts around $1050$ms when EHOs appear. There is a sporadic ELM at $\sim 1500$ ms and then the EHOs reappear strongly. This can also be seen in part (b), which shows the "quiet" $D_{\alpha}$ profile when EHOs are present. More details about the experiment can be found in Ref~\cite{chen2016rotational}.}
\label{fig:confexp}
\end{figure}

\section{\label{sec: experimentalsetup} Experimental Setup and Inputs}
\begin{figure}
\includegraphics[width=0.49\textwidth]{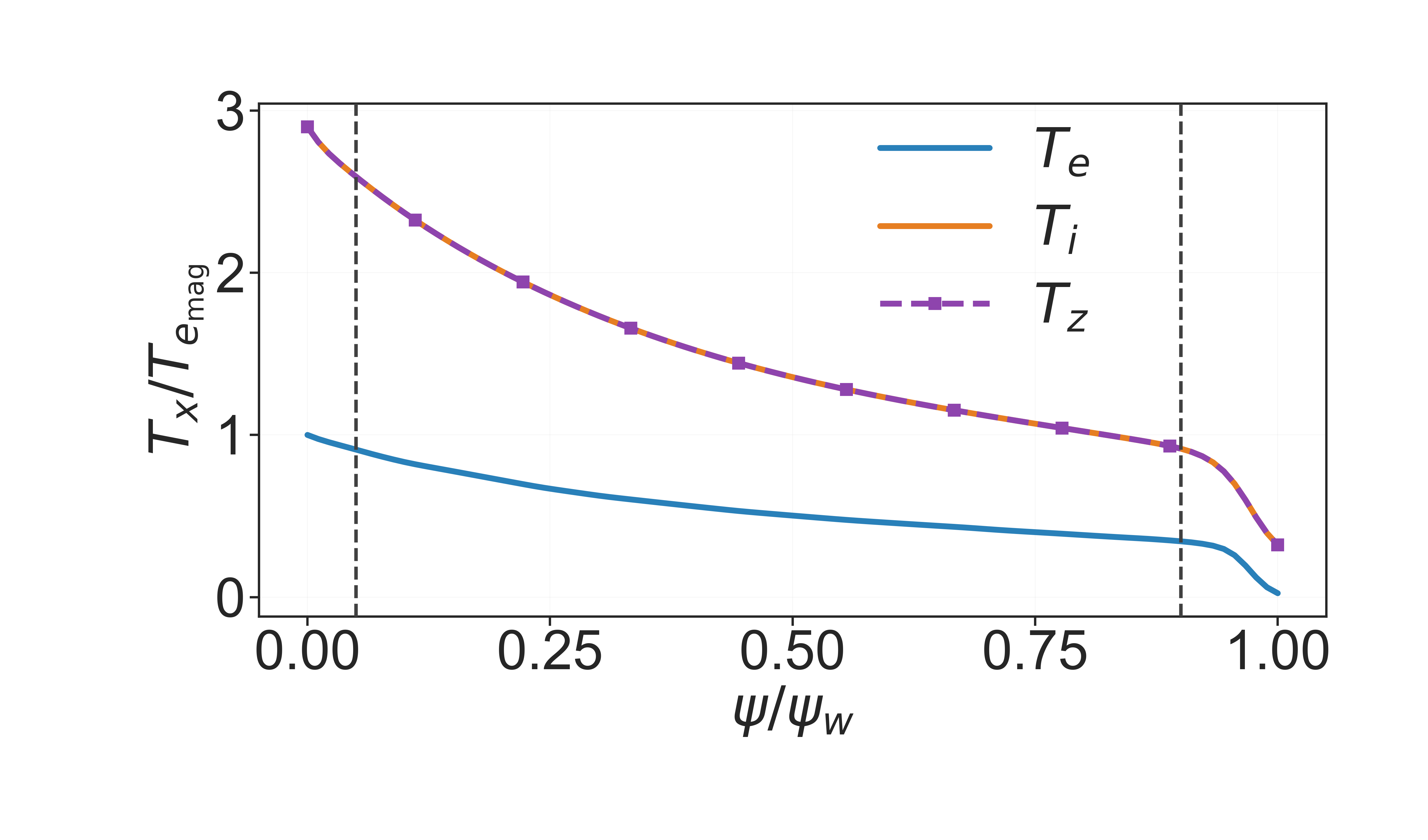}
\includegraphics[width=0.49\textwidth]{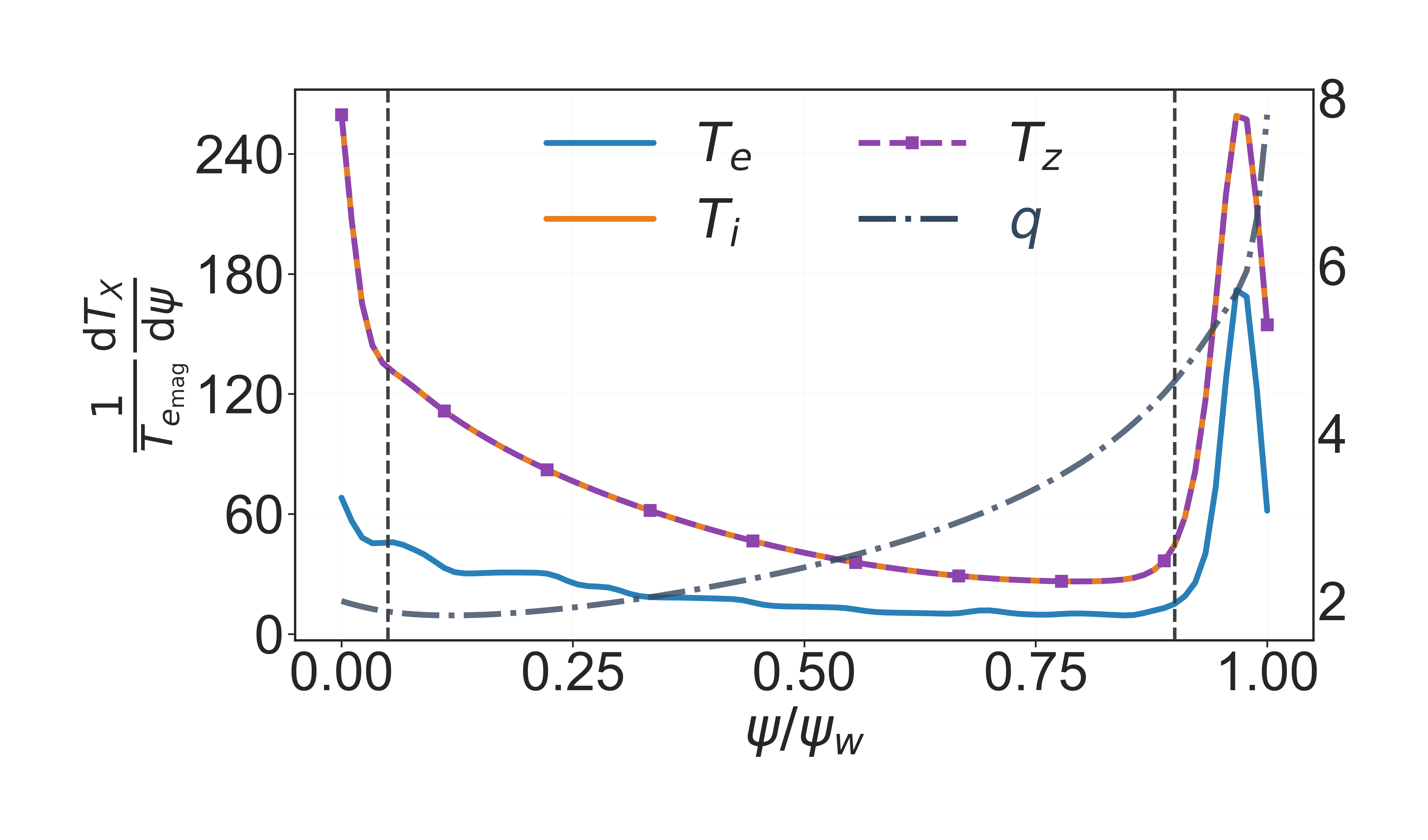}
\centering
\caption{\label{fig:tempprofile} (Left) The original temperature profile obtained from 
the experiment. The ion temperature ($\sim 10.24$ keV) is higher than the electron temperature ($\sim 3.54$ keV) due to NBI 
injection. The impurity ion temperature (impurity taken here to be Carbon and its temperature is denoted by $T_z$ in the above figure) is the same as the main ion temperature for the reasons described in Sec~\ref{sec:impstudy}. All the quantities are normalized by on--
axis electron temperature. The grey dotted lines mark the simulation domain $\psi/\psi_w \in [0.05,0.9]$ where $\psi$ is the flux surface. (Right) We plot the temperature gradient normalized by on-axis electron temperature as a function of $\psi/\psi_w$.}
\end{figure}

\noindent We briefly describe the experimental conditions for the DIII-D QH-mode discharge \#157102 that we use for our gyrokinetic simulations, the results of which are presented in this manuscript. More details are available in \cite{chen2016rotational}. Figure~\ref{fig:confexp} shows the time evolution of relevant discharge parameters and the magnetic fluctuation spectrogram, which indicates that the QH-mode with strong EHO activity is sustained until $\sim 2950$ ms. This discharge has a torque ramp down starting at $\sim 3000$ ms for other studies \cite{chen2016rotational} without strong EHO activity. The discharge transitions from L-mode to H-mode around $980$ ms as evidenced by an increase in the pedestal electron pressure shown in Figure~\ref{fig:confexp}c. The discharge then transitions into QH-mode around $1040$ ms when the EHOs with fundamental harmonic at $\sim 10$ kHz appear in the magnetic fluctuation spectrogram as shown in Figure ~\ref{fig:confexp}d. With the appearance of the EHOs, the D$\alpha$ light intensity (Figure~\ref{fig:confexp}b) increases and is correlated with the integrated RMS amplitude (Figure~\ref{fig:confexp}e) of the EHOs calculated from the spectrogram shown in Fig~\ref{fig:confexp}d. NBI power of $5.5-6$ MW and NBI torque of $4.5-5.5$ Nm are injected (see Figure~\ref{fig:confexp}a). Pedestal electron pressure of $\sim 2$ kPa and chord-averaged electron density of $\sim 1.7 \times 10^{13}/\text{cm}^3$ (Figure~\ref{fig:confexp}g) is maintained in the QH-mode phase. This discharge is highly shaped, near balanced double null, with elongation $\kappa \sim 1.9$, average triangularity, $\delta \sim 0.58$. This discharge has a toroidal magnetic field $B_T \sim 2.06$ T and the plasma current $I_P \sim 1.1$MA. The edge safety factor $q_{95} \sim 5.4$ and normalized Beta, $\beta_N\sim 1.6$ (Figure~\ref{fig:confexp}f). This discharge is produced by counter-injecting NBI power in a direction opposite to the plasma current. We take all our plasma profiles (temperature profile and density profile as shown in Fig~\ref{fig:tempprofile} and Fig~\ref{fig:denprofile}) around $2420$ms as shown in the grey box in Fig~\ref{fig:confexp}. These profiles serve as the input conditions for our gyrokinetic simulations, enabling a direct study of core transport in experimentally relevant QH-mode plasmas. In the following sections, we investigate the sensitivity of microturbulence and transport to variations in the ion and electron temperatures, starting with a systematic reduction in ion temperature while holding electron temperature fixed.

\section{\label{sec: simDion} Simulation of Microinstabilities with Decreasing Ion temperature}

\subsection{\label{sec: lin} Linear Regime}
\begin{figure}
\includegraphics[width=0.49\textwidth]{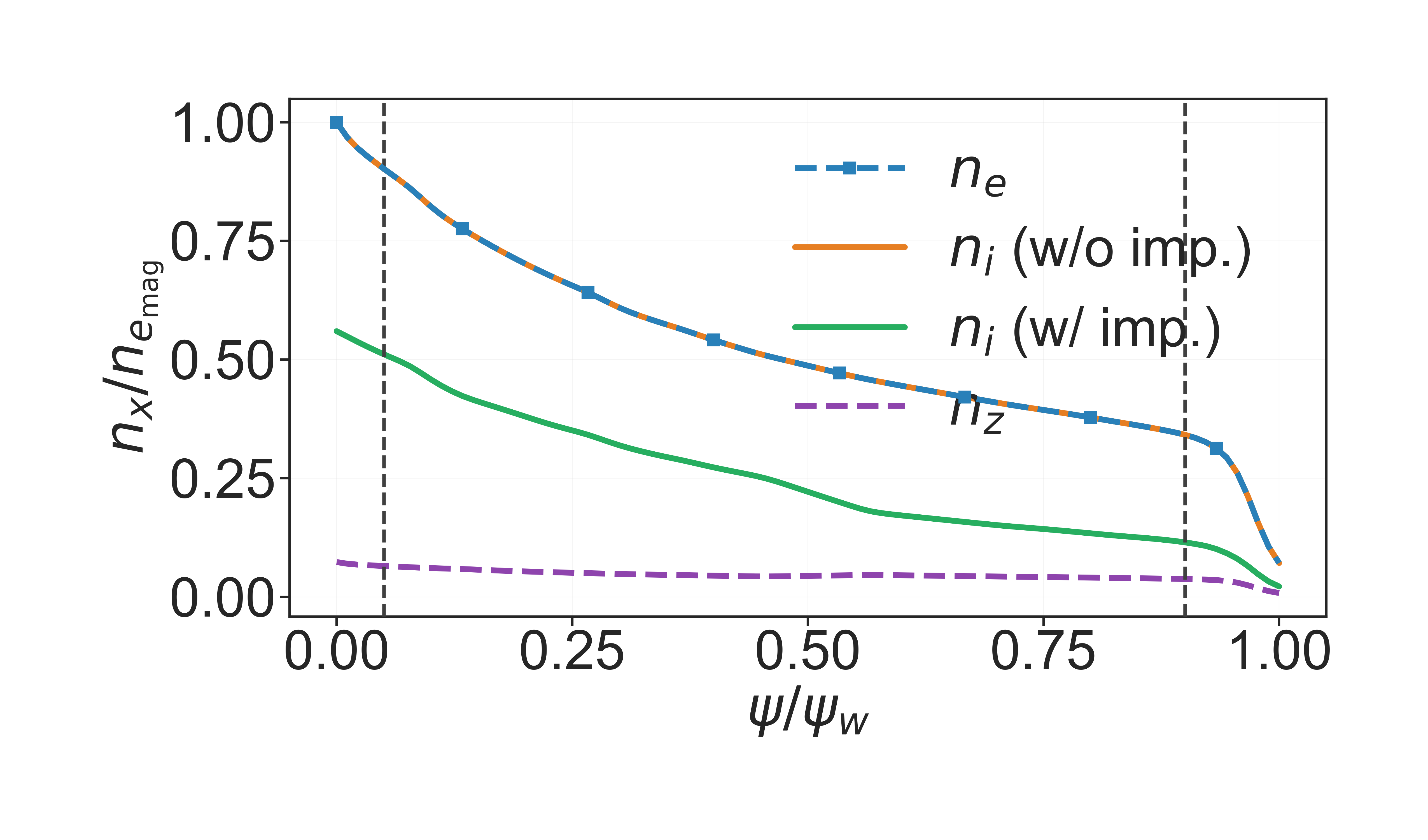}
\includegraphics[width=0.49\textwidth]{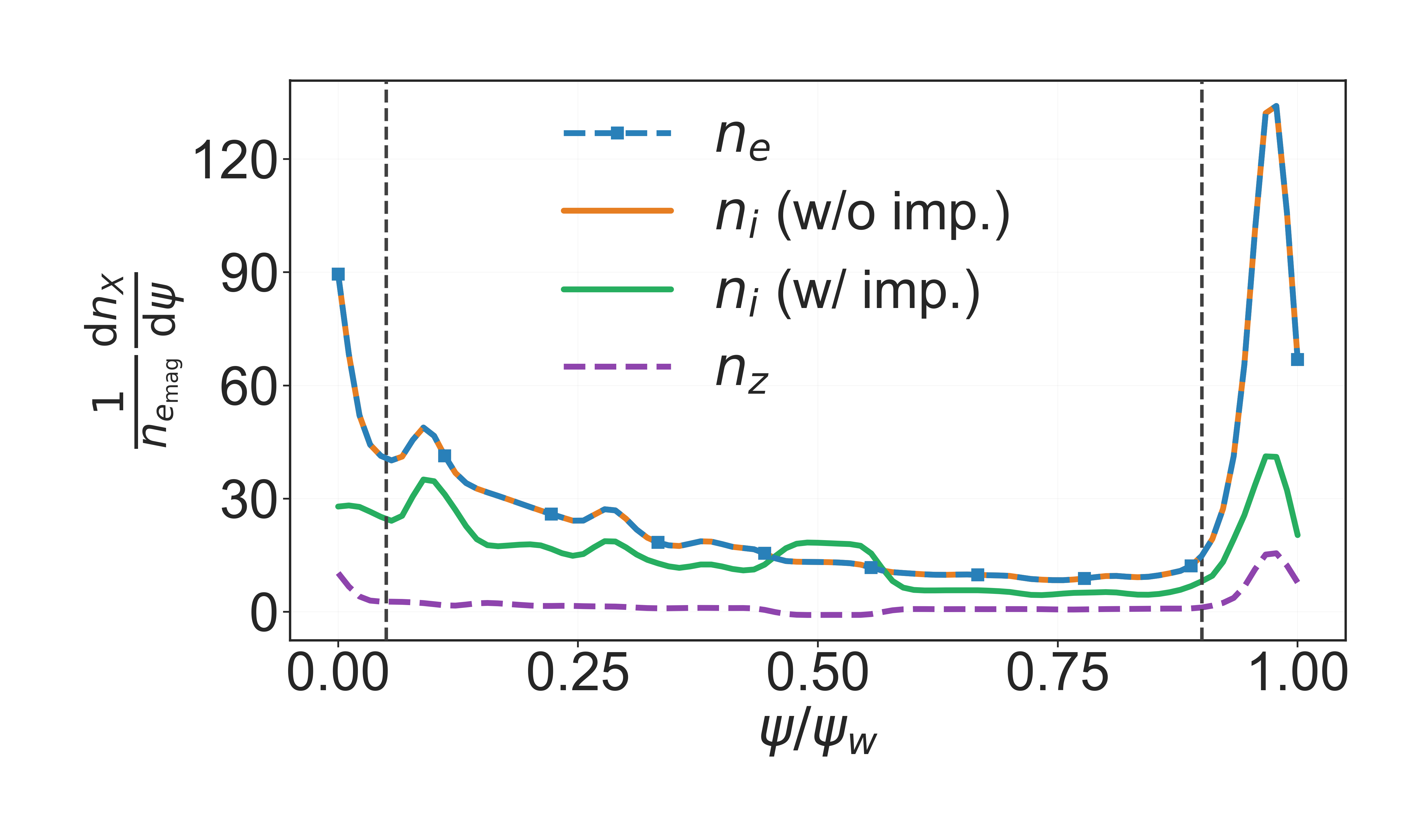}
\centering
\caption{\label{fig:denprofile} (Left) Experimental electron density profile used in the simulations, normalized to the on-axis electron density, $n_{e_{\text{mag}}}$. The grey dotted lines mark the simulation domain $\psi/\psi_w \in [0.05,0.9]$. We are using deuterium plasma in our simulation. The electron density profile is the same whether the impurities (impurity taken here to be Carbon and its density is denoted by $n_z$ in the above figure) are present or absent. Due to the quasineutrality condition in Eq.~\ref{eq:quasineutrality}, the ion density differs when the impurity is present or absent. When no impurities are present, then $n_i$(w/o imp) = $n_e$, shown in blue and orange colors. When the impurities are absent, $n_i$(w/ imp)= $n_e$- $6\times n_z$, shown in green color in the above figure. (Right) We plot the density gradient normalized by on-axis electron density as a function of $\psi/\psi_w$.
}

\end{figure}

\noindent In this section, we carry out four different linear simulations with decreasing ion temperature without considering any impurities in these cases. The original temperature profile obtained in the discharge is shown in Fig~\ref{fig:tempprofile}. $T_i$ and $T_e$ in the core are $10.24$ keV and $3.54$ keV respectively. $T_i$ is higher due to NBI power injection, which primarily heats the ions,  along with the low collisionality of plasma ($\nu_*\sim0.2$). We perform simulations in the core region ($\psi/\psi_w \in [0.05,0.9]$) marked with grey dotted lines in Fig~\ref{fig:tempprofile}, where $\psi_w= 0.036$ is the $\psi$ at separatrix. We plot the density profile in Fig~\ref{fig:denprofile}. The on-axis electron density is $n_e=3.14\times 10^{19}$ m$^{-3}$, and the ion density is $n_i= 1.33\times 10^{19}$ m$^{-3}$. We use deuterium plasma in our simulation ($n_e=n_i$ (w/o imp.) in Fig~\ref{fig:denprofile}). We plot the safety factor($q$) in Fig~\ref{fig:tempprofile} on the right-hand side with the temperature gradients.

\noindent Our simulations fully account for the kinetic effects of the electrons. We first study the linear regime for the case $T_i/T_e=2.90$ (original case). The major radius is $R_0=1.75m$; the on-axis magnetic field is $B_T=1.96T$. After the convergence test, we use $32$ parallel grid points, $96$ radial grid points, $1500$ poloidal grid points, $50$ ions per cell, and $\Delta t= 0.01 R_0/C_s$ where $C_s/R_0=23.53 \times 10^4$sec$^{-1}$ and $C_s=\sqrt{T_e/m_i}$ is the ion acoustic speed. 
Table~\ref{tab:linion1} presents the results for decreasing ion temperature at the magnetic axis while keeping the electron temperature fixed. The ion temperature is decreased from $10.24$ keV to $1.22$ keV. We focus on four $T_i/T_e$ temperature ratios: $T_i/T_e=\{2.90,1.45,1.0,1/2.90\}$. Following a convergence test, we set the poloidal grid to 2500 points for all $T_i/T_e$ cases except the original, keeping all other simulation parameters unchanged. Further, only for $T_i/T_e=1/2.90$, we chose $\Delta t=0.005 R_0/C_s$ as the time step.

In Fig~\ref{fig:zfmmodestructnlo} (a), we show the linear phase of the normalized electrostatic perturbed potential on the poloidal plane of $\zeta=0$ at $t=37.50R_0/C_s$ for the original case. The poloidal and toroidal mode numbers at the location where the eigenmode peaks ($\psi\sim 0.29\psi_w$) are m$\sim 35$ and n$\sim 17$, respectively. The growth rate for this case is $\gamma=0.30 C_s/R_0$.

\begin{figure}
\includegraphics[width=0.48\textwidth]{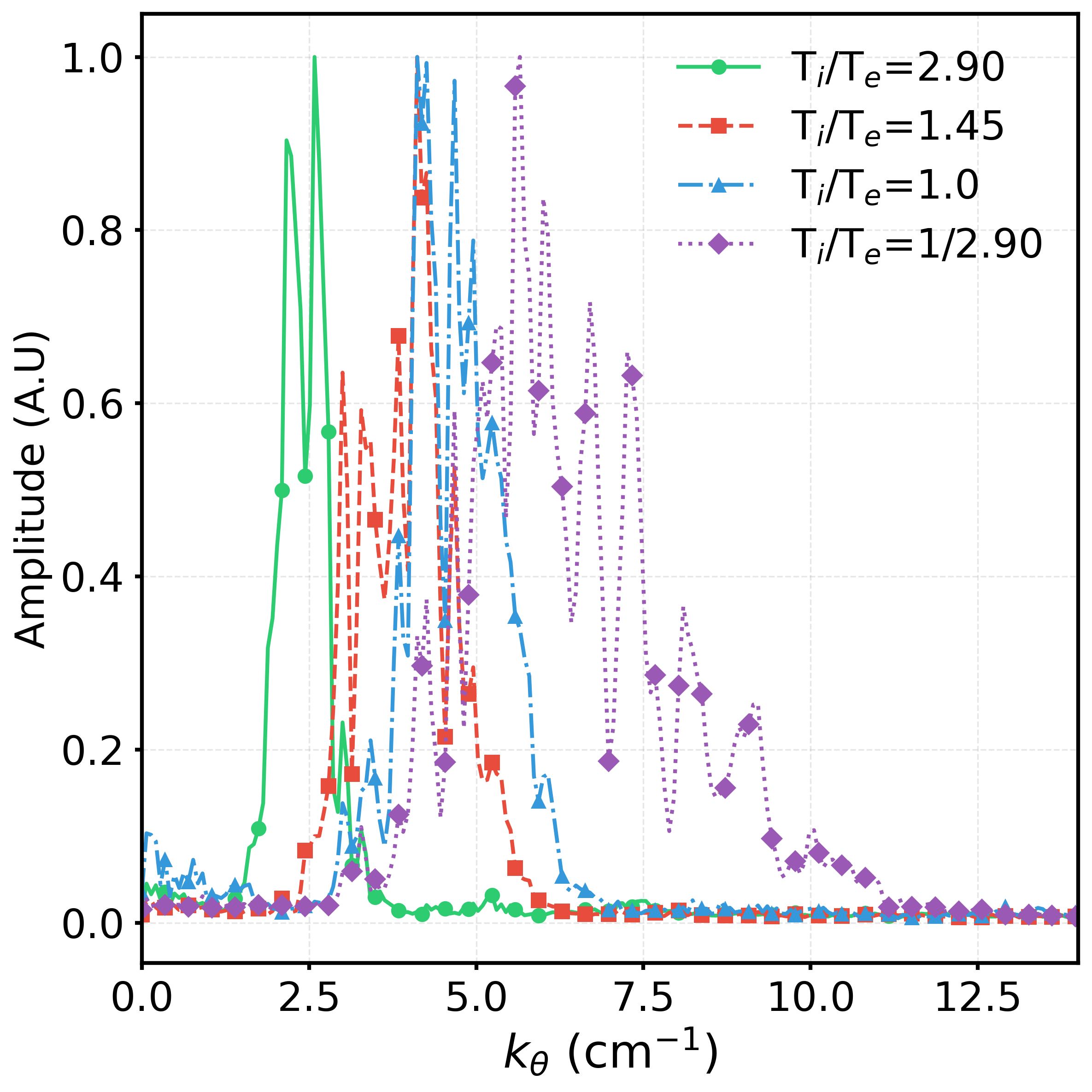}
\includegraphics[width=0.48\textwidth]{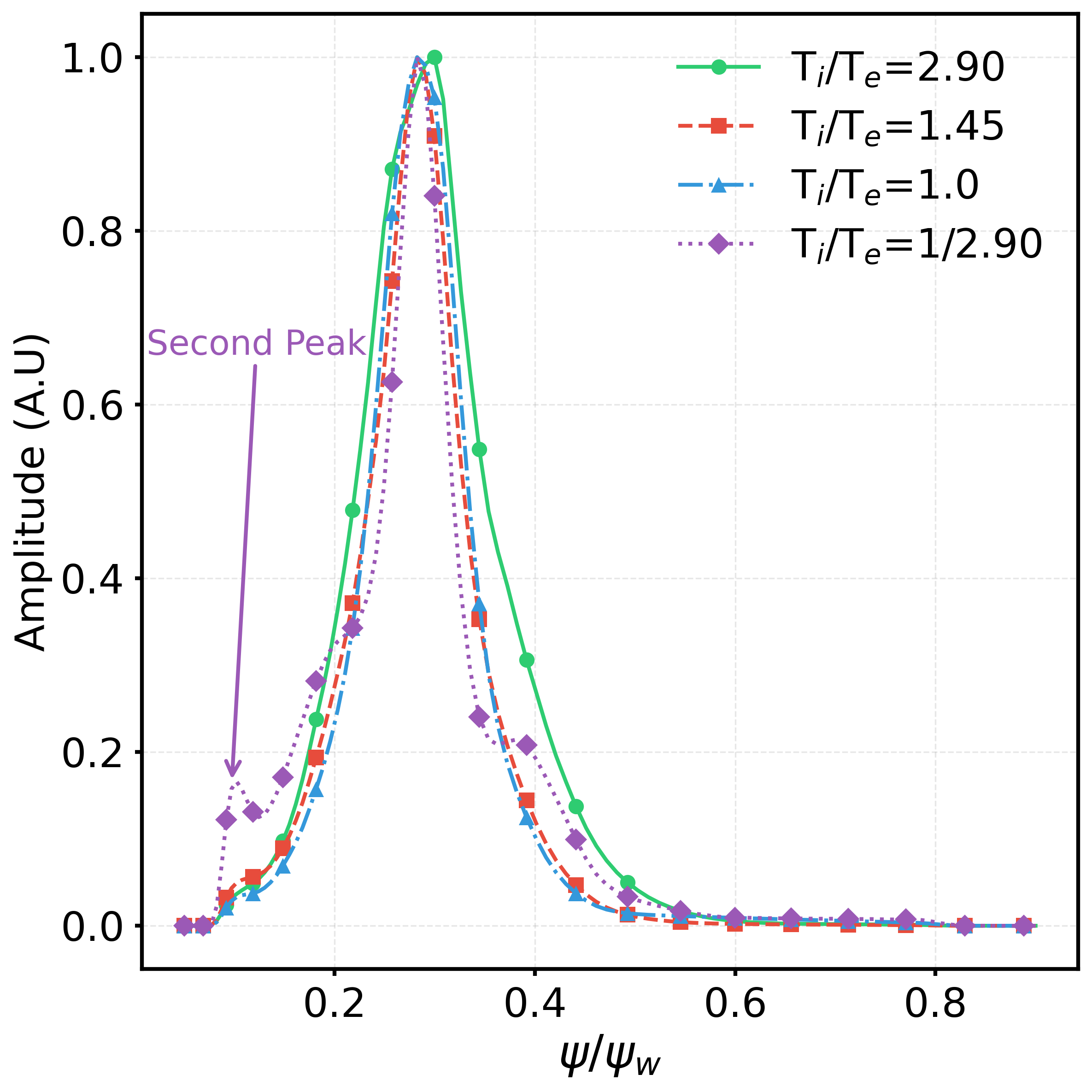}
\centering
\caption{\label{fig:mnstructureion} (Left)  Poloidal spectrum in the linear regime at $t=25R_0/C_s$ for the four $T_i/T_e$ ratios, where the maximum amplitude is normalized to 1. The poloidal mode number increases as we decrease the $T_i/T_e$ ratio, with electron temperature fixed while only the ion temperature is varied. (Right) We plot the radial profile in the linear regime at $t=25R_0/C_s$ for the four $T_i/T_e$ ratios, also normalized to a maximum amplitude of 1. All modes peak at the same location of $\psi/\psi_w=0.29$. Additionally, we observe a secondary peak around $\psi/\psi_w=0.1$, identified as an ITG mode for $T_i/T_e=2.90$ and a TEM mode for $T_i/T_e=1.0, 1/2.90$. For $T_i/T_e=1.45$, the mode diamagnetic frequency is zero at that location.}

\end{figure}

\noindent Table~\ref{tab:linion1} presents the results for the linear regime as the ratio $T_i/T_e$ decreases from $2.90$ to $1/2.90$. First, the poloidal and toroidal mode numbers increase as we reduce the $T_i/T_e$ ratio. The growth rate increases from $\gamma= 0.30C_s/R_0$ at the highest $T_i/T_e$ ratio to $\gamma= 0.60C_s/R_0$ at the lowest. In the original case, corresponding to the highest $T_i/T_e$ ratio, as well as for $T_i/T_e=1.45$, the mode frequency is directed along the ion diamagnetic direction, thereby classifying the instability as ITG according to the GTC convention. When we decrease the temperature ratio to $T_i/T_e=1$ and further to $T_i/T_e= 1/2.90$, the mode frequency changes to the electron diamagnetic direction. By GTC convention, this mode is a TEM mode. Hence, we can say that as we decrease the $T_i/T_e$ ratio, ITG mode is relatively stabilized, whereas TEM is destabilized\cite{casati2008temperature,mckee2014turbulence,sommer2015transport}. We further find that all the modes peak at the same location, as shown in the right side of Fig~\ref{fig:mnstructureion}. Additionally, we see a subdominant mode at $\psi/\psi_w=0.1$ in the radial profile in Fig~\ref{fig:mnstructureion}(Right). We verify this subdominant to have the frequency in the direction of ion (ITG) in the case of $T_i/T_e=2.90, 1.45$ and the direction of electron(TEM) for the cases $T_i/T_e=1.0$ and $T_i/T_e=1/2.90$.

\begin{table}

\begin{threeparttable}
\begin{tabular*}{\textwidth}{@{}l*{4}{@{\extracolsep{0pt plus 12pt}}l}}
\hline
$T_i$ & $10.24$ keV & $7.06$ keV & $3.54$ keV & $1.22$ keV \\
\hline
$T_i/T_e$ & $2.90$ (Original case) & $1.45$ & $1$ & $1/2.90$ \\
\hline
m $(k_\theta [cm^{-1}])$ & $35 (2.4)$ & $59 (4.1)$ & $59 (4.1)$ & $81 (5.7)$ \\
n & $17$ & $30$ & $30$ & $41$ \\
$\gamma (R_0/C_s$ units) & $0.30$ & $0.46$ & $0.51$ & $0.60$ \\
Mode & ITG & ITG & TEM & TEM \\
% Width (r/a units) & $0.059$ & $0.067$ & $0.096$ & $0.126$ \\
Width (r/a units) & $0.06$ & $0.07$ & $0.10$ & $0.13$ \\
\hline
\end{tabular*}
\caption{\label{tab:linion1} Comparison of the various quantities in the linear simulation for the case when we decrease the ion temperature, keeping the electron temperature fixed at $3.54$ keV.}
 % \begin{tablenotes}
 %    \item[*] Inconclusive Case.
 %  \end{tablenotes}
\end{threeparttable}
\end{table}

\subsection{\label{subsec:nldi} Non Linear Regime}

\begin{figure}
    \centering
    \begin{overpic}[width=0.32\textwidth]{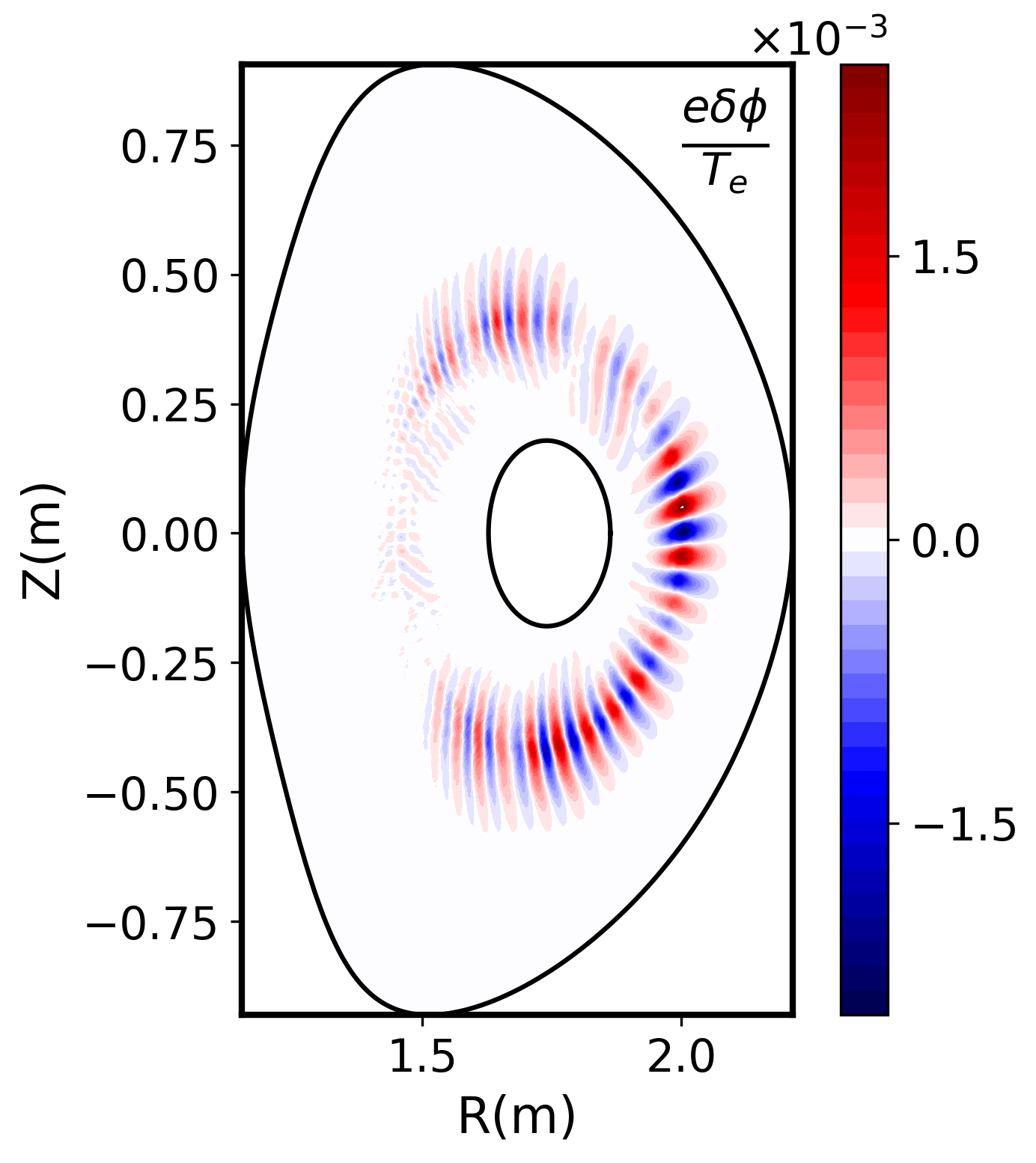}
        \put(60,20){(a)}
    \end{overpic}
    \begin{overpic}[width=0.32\textwidth]{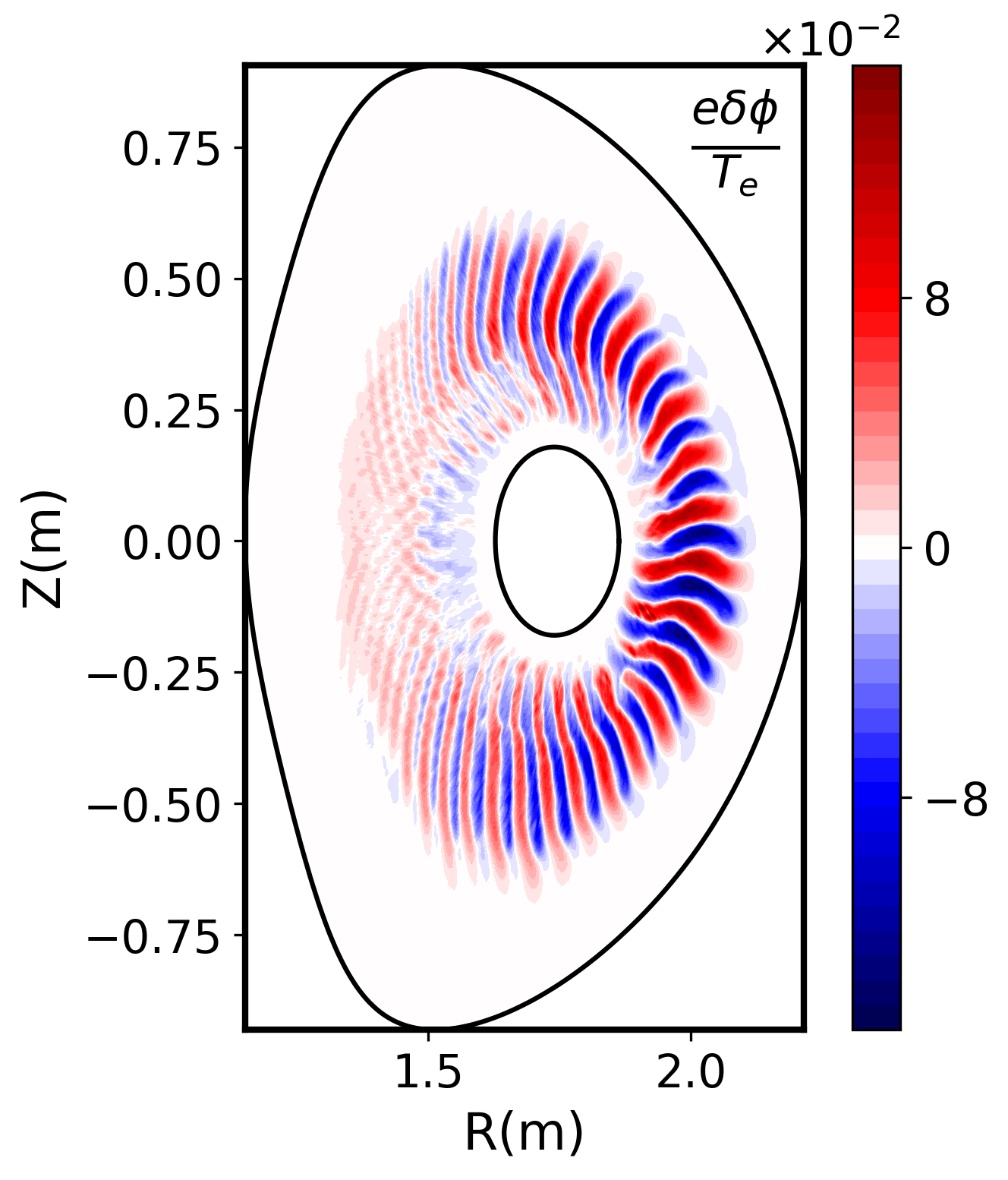}
        \put(60,20){(b)}
    \end{overpic}
    \begin{overpic}[width=0.32\textwidth]{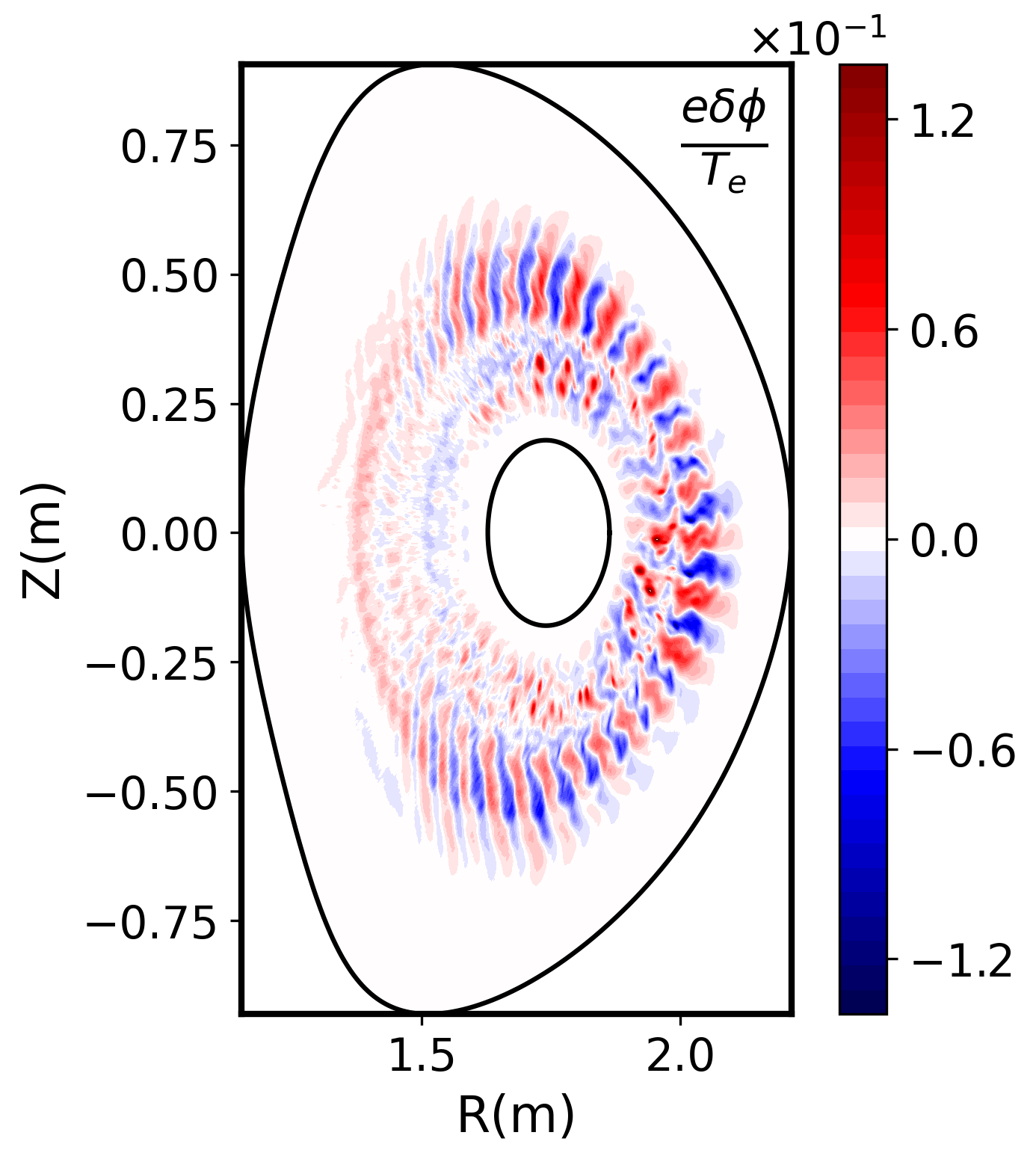}
        \put(60,20){(c)}
    \end{overpic}
    \caption{\label{fig:zfmmodestructnlo} (a) Contour plots of the electrostatic perturbed potential in the linear phase on $\zeta=0$ poloidal plane at $t=37.5R_0/C_s$,(b) nonlinear phase without ZF, (c) nonlinear phase with ZF, both at $t=57.50 R_0/C_s$. The black curves indicate the inner and outer simulation boundaries.}
\end{figure}

\noindent We now study the nonlinear regime for this case. In Fig~\ref{fig:zfmmodestructnlo} (b-c), we show the effect of ZFs on the electrostatic potential in the nonlinear regime for the original case. In the absence of ZFs, we see that the linear mode structure spreads radially from the linear eigenmode. With the ZFs, the mode structure breaks. We can measure the radial size of the eddies in the presence of ZF by calculating the correlation length defined as\cite{xiao2009turbulent}

\begin{eqnarray}
    C_{r\theta}(\Delta r, \Delta \theta) = \frac{\langle \delta \phi (r + \Delta r, \theta + \Delta \theta) \delta \phi (r, \theta) \rangle}{\sqrt{\langle \delta \phi^2 (r + \Delta r, \theta + \Delta \theta) \delta \phi^2 (r, \theta) \rangle}},
    \label{eq:correlation}
\end{eqnarray}

Here $\langle ...\rangle$ denotes the flux surface averaging, $\Delta r$ and $\Delta \theta$ denote the radial and poloidal separation between the points, respectively. We calculate the $1$D correlation function, $C_r(\Delta r)$ by taking the maximal value along the ridge of the above $2$D correlation function, which exhibits the Gaussian decay: $C_r(\Delta r)\sim \exp(-\Delta r/ L_r)$ where $L_r$ is the radial correlation length. We plot $L_r$ in Fig~\ref{fig:correlation}. We see that as the $T_i/T_e$ ratio is decreased for this case, the radial correlation length decreases which signifies that the size of eddies decreases as the $T_i$ is decreased.

\begin{figure}
\includegraphics[width=0.48\textwidth]{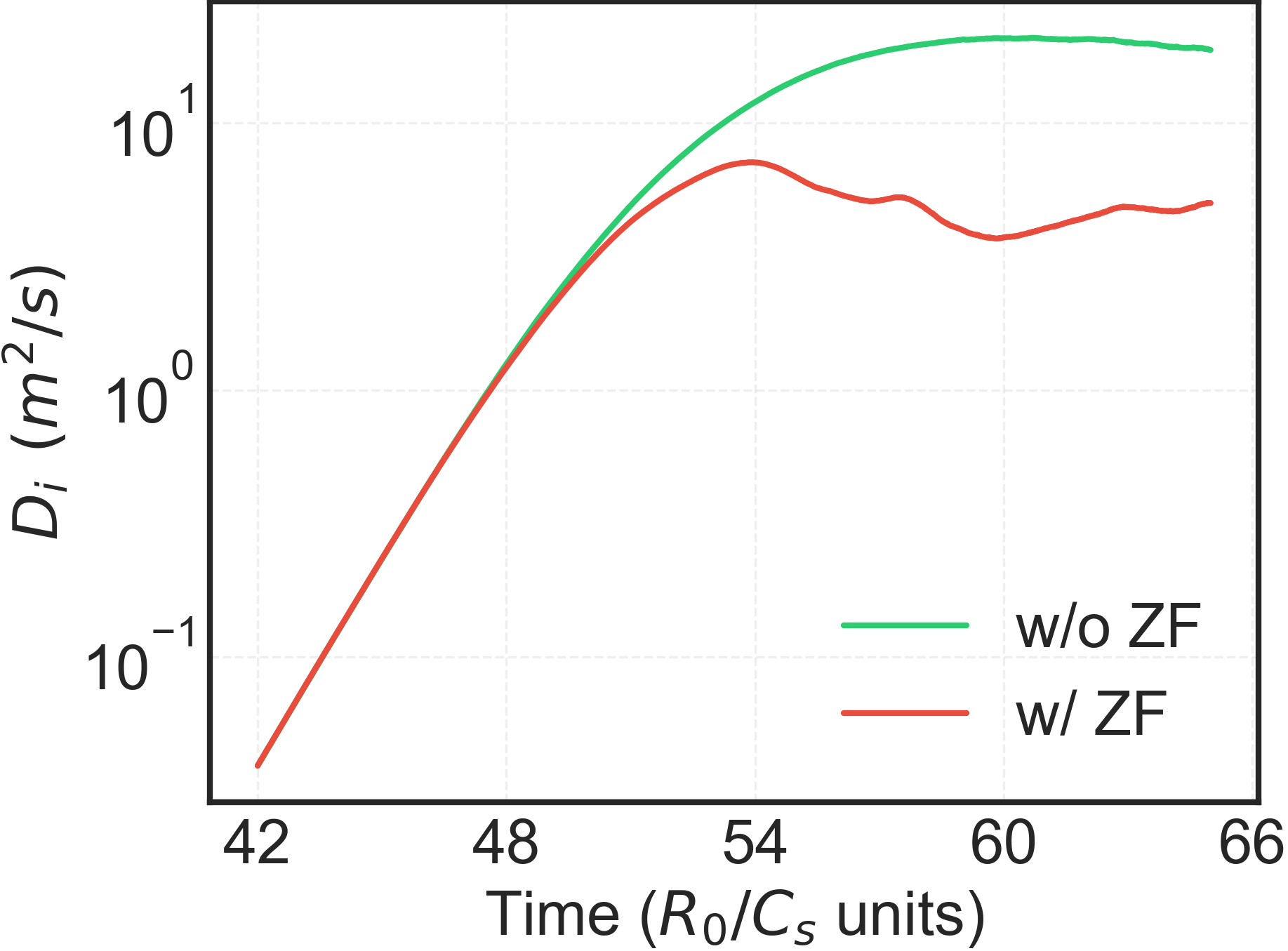}
\includegraphics[width=0.48\textwidth]{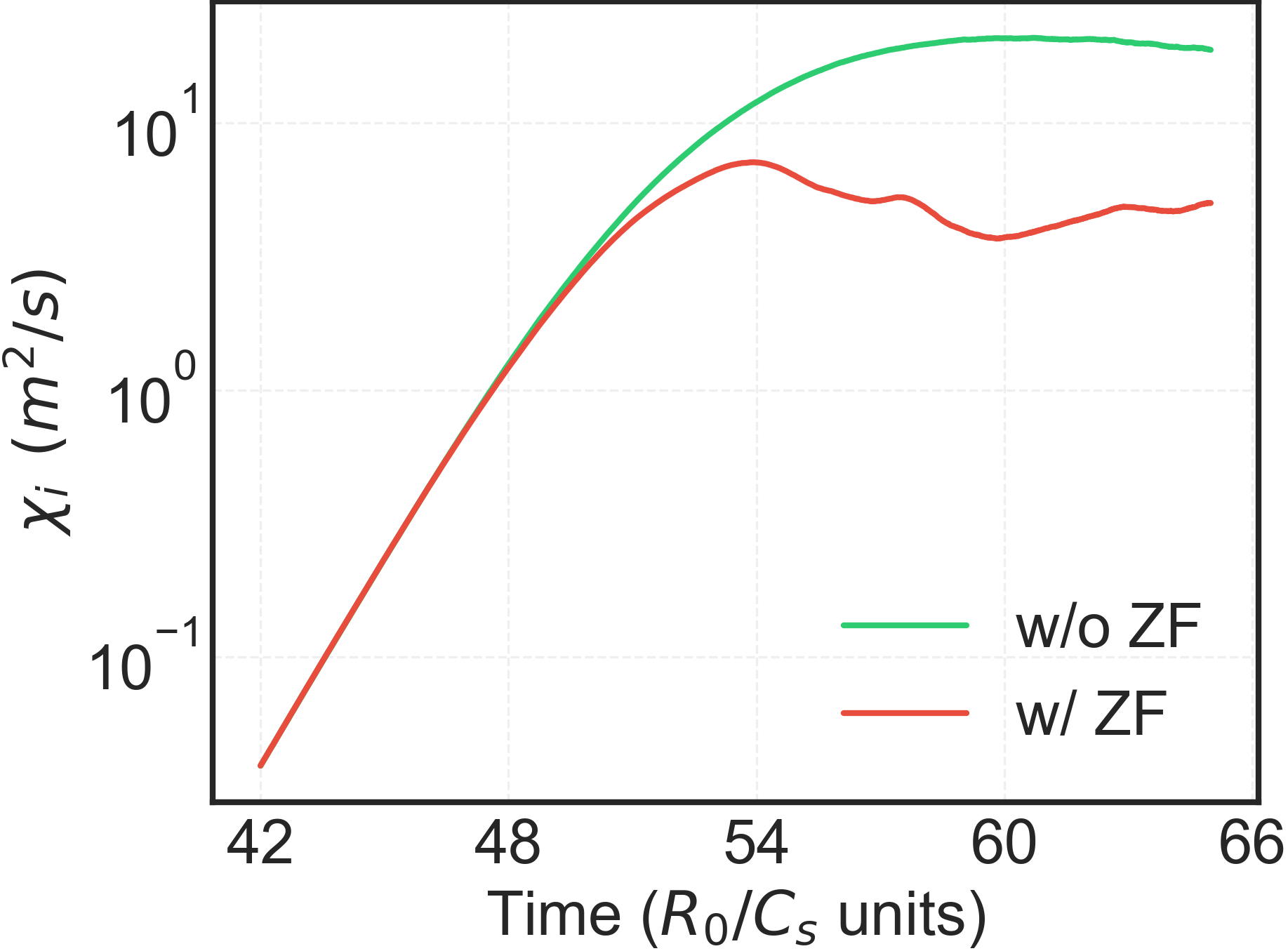}
\centering
\caption{\label{fig:zfo1} Effect of ZFs on the heat and particle diffusivities. In the presence of ZFs, the ion particle diffusivities are reduced by $5.82\times$ (Left), and the ion heat diffusivities are reduced by $2.96\times$ (Right) for the case $T_i/T_e=2.90$. A similar reduction is found for other temperature ratios in electron heat and particle diffusivity cases.}

\end{figure}

In Fig~\ref{fig:zfo1}, we show the effect of ZFs on the transport quantities. We plot the ion particle diffusivity ($D_i$) and the ion heat diffusivity ($\chi_i$). We find that ZFs lead to a reduction in the transport quantities by a factor of $5.82\times$ and $2.96\times$. We see a similar decrease in the electron transport quantities. We also find that ZFs reduce the transport quantities for other cases of $T_i/T_e$. Further, we find that the reduction due to ZFs is highest when $T_i/T_e=2.90$, i.e., when the ion temperature is highest, and it reduces as we decrease it.

\section{\label{sec: simele} Simulation of Microinstabilities with Increasing Electron Temperature}

\subsection{\label{subsec:linei} Linear Regime}
\noindent We now change the temperature ratio by changing the electron temperature, keeping the ion temperature fixed. Again, we do not consider the effect of impurity ions in this case. We used the same simulation parameters for this case as in our previous study of decreasing the ion temperature at fixed electron temperature. The main results for the linear regime for this case are presented in Table~\ref{tab:linele1}. We note differences from the previous case, where we decreased the ion temperature. First, the poloidal and toroidal mode numbers decrease as we increase the electron temperature. We expect this result since $k_{\theta}=m/r \sim 1/\rho_s\sim 1/\sqrt{T_e}$\cite{horton1999drift}. Here, $\rho_s= C_s/\omega_{ci}$ is the ion gyroradius. Here, $\omega_{ci}= eB/m_i$ is the ion gyrofrequency. Since the electron temperature increases, we expect $k_{\theta}$ to decrease. We further see that the growth rate increases as the electron temperature increases\cite{sommer2015transport}. We observe that the growth rate (in $R_0/C_s$ units) remains the same for a given $T_i/T_e$ ratio, whether we decrease the ion temperature or increase the electron temperature \cite{sommer2015transport}. However, in physical units ($\gamma C_s/R_0)$, the growth rate is higher when increasing the electron temperature~\cite{casati2008temperature}. This is because $C_s\sim \sqrt{T_e}$, while $R_0$ remains constant, leading to a growth rate, in the case of increasing $T_e$, higher by a factor of $\sqrt{T_e/T_{e_o}}$, where $T_{e_o}= 3.53$keV is the original electron temperature. Further, we find that for the case $T_i/T_e={2.90, 1.45}$, the mode propagates in ion diamagnetic direction (ITG), and for the case $T_i/T_e={1.0,1/2.90}$, the mode propagates in electron diamagnetic direction (TEM)\cite{sommer2015transport,mckee2014turbulence}.

\subsection{\label{subsec:nlei} Non-Linear Regime}
\noindent We now study the nonlinear regime for the present case. We observe a similar trend in the reduction of transport quantities due to ZFs when increasing the electron temperature, as seen previously when decreasing the ion temperature for all temperature ratios. Further, we see that the mode structure eddies in this case show a larger width than in the case with decreasing ion temperature. We observe this in Fig~\ref{fig:correlation} where we have plotted the radial correlation length calculated using Eq~\ref{eq:correlation}. We can explain this observation by noticing that the radial width of eddies is proportional to $\rho_s$, which further depends on the electron temperature\cite{horton1999drift}. Hence, increasing $T_e$ should increase the radial width of eddies. 

\section{\label{sec:comparetwostudies} Comparison of the two studies}

This section presents a comparative analysis of the two preceding studies. We examine two approaches to modify the $T_i/T_e$ temperature ratio: reducing the $T_i$ and increasing the $T_e$. The impact on poloidal and toroidal mode numbers reveals that in both scenarios—decreasing $T_i$ and increasing $T_e$—the mode numbers decrease with declining $T_i/T_e$ ratio \cite{horton1999drift}. For identical values of $T_i/T_e$, the growth rate in physical units is more pronounced when increasing $T_e$ compared to decreasing $T_i$ as explained in Sec~\ref{subsec:linei}. Additionally, both approaches demonstrate ITG stabilization and TEM destabilization \cite{casati2008temperature,he2024ti}.

\begin{figure}
\includegraphics[width=0.48\textwidth]{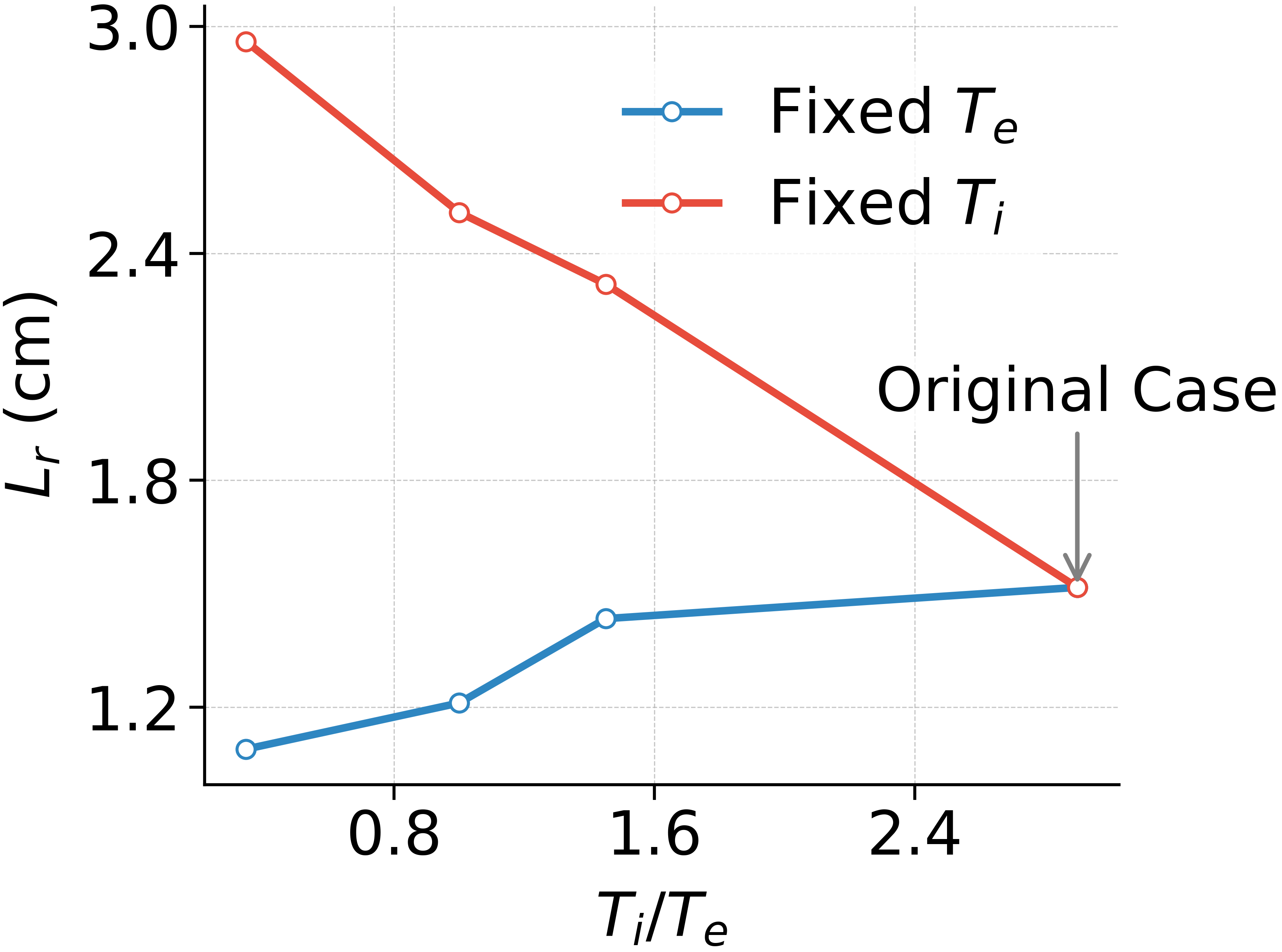}
\centering
\caption{\label{fig:correlation} Variation of correlation length, $L_r$ in $cm$ as a function of $T_i/T_e$ for two distinct cases. In the first case (Fixed $T_e$), as $T_i$ is decreased, the correlation length does not change significantly, indicating that the radial size of eddies does not change significantly. In the second case (Fixed $T_i$), we observe that the correlation length increases with increasing $T_e$, demonstrating that the radial size of eddies increases with increasing electron temperature.}

\end{figure}

\begin{table}

\begin{threeparttable}
\begin{tabular*}{\textwidth}{@{}l*{4}{@{\extracolsep{0pt plus 12pt}}l}}
\hline
$T_e$ & $3.53$ keV & $5.13$ keV & $10.24$ keV & $29.59$ keV \\
\hline
$T_i$/$T_e$ & $2.90$ (Original case) & $1.45$ & $1$ & $1/2.90$ \\
\hline
m $(k_\theta [cm^{-1}])$ & $35 (2.4)$ & $41 (2.9)$ & $33 (2.3)$ & $28 (2.0)$ \\
n & $17$ & $20$ & $16$ & $14$ \\
$\gamma (R_0/C_s$ units) & $0.30$ & $0.43$ & $0.51$ & $0.57$ \\
Mode & ITG & ITG & TEM & TEM \\
% Width (r/a units) & $0.059$ & $0.096$ & $0.134$ & $0.171$ \\
Width (r/a units) & $0.06$ & $0.10$ & $0.13$ & $0.17$ \\
\hline
\end{tabular*}
\caption{\label{tab:linele1} Comparison of the various quantities in the linear simulation for the case when the electron temperature is increased, keeping the ion temperature fixed at $10.24$ keV.}
\end{threeparttable}
\end{table}

In the nonlinear regime, however, the transport response differs markedly depending on the heating method, as illustrated in Tables~\ref{tab:satvaliond} and~\ref{tab:satvalele}. In physical units, our analysis reveals that reducing the ion temperature from the original case ($T_i/T_e=2.90$) increases the saturation level of ion heat diffusivity modestly. In contrast, elevated electron temperatures lead to a substantially higher transport enhancement of $\sim 31$-fold for the same final $T_i/T_e$ ratio. These findings are consistent with observations in previous experimental studies\cite{mckee2014turbulence,petty1999dependence}.

\begin{figure}
	
\includegraphics[width=0.45\textwidth]{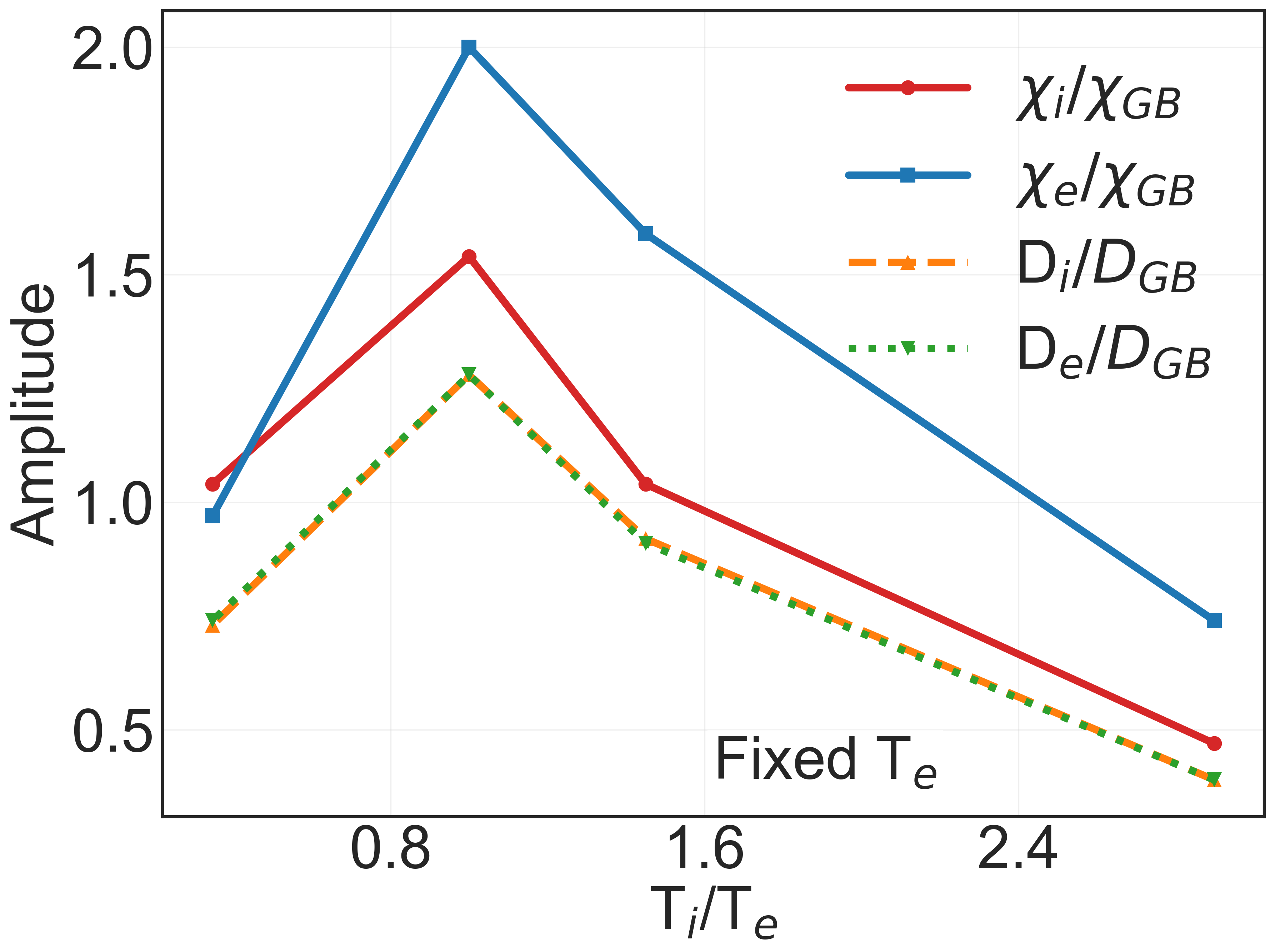}
\includegraphics[width=0.45\textwidth]{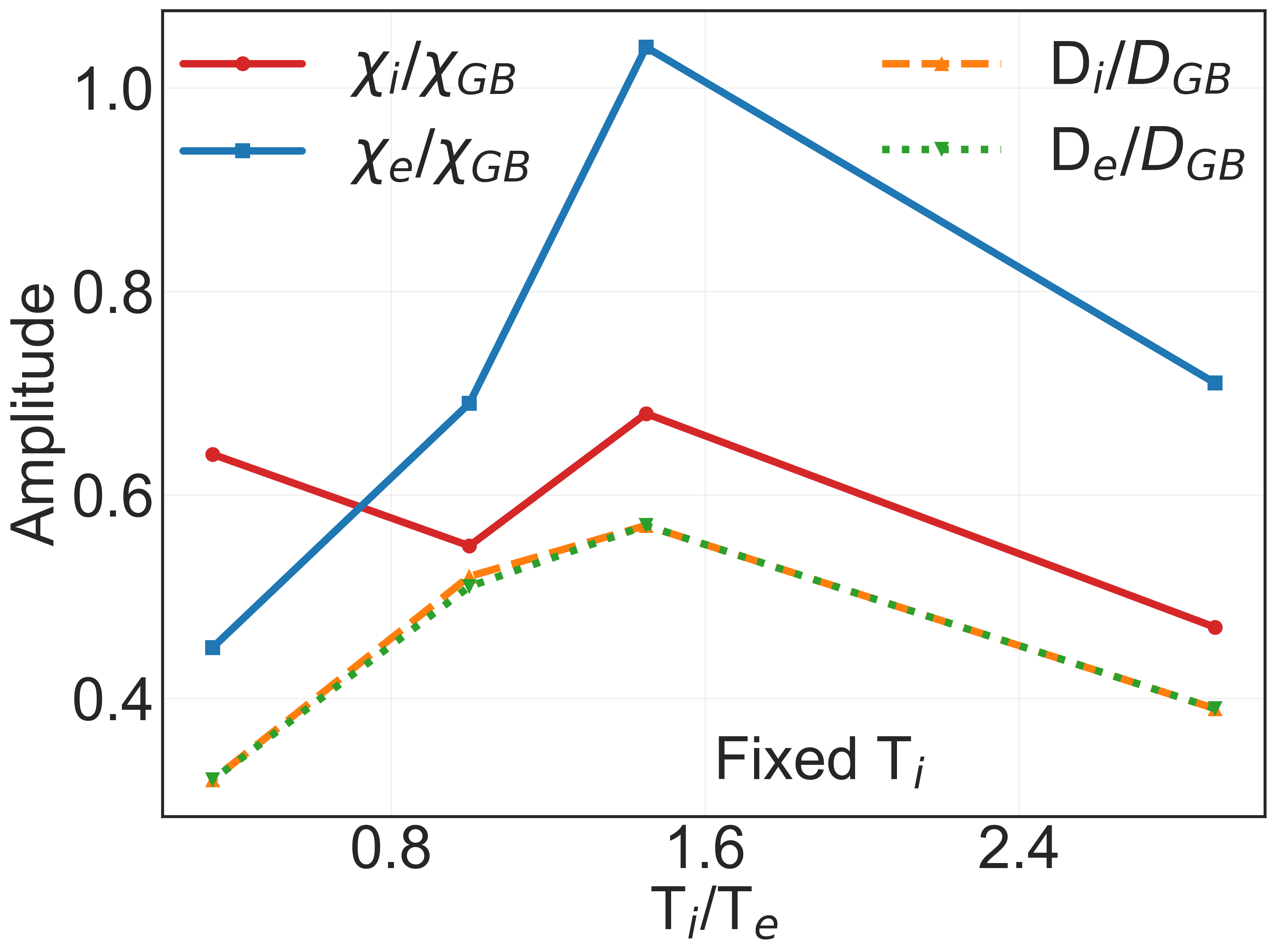}
\centering
\caption{\label{fig:tabcmpfig} Comparison of the normalized transport quantities as a function of $T_i/T_e$ between the two studies presented above. (Left) Decreasing ion temperature at fixed electron temperature. (Right) Increasing electron temperature at fixed ion temperature. In both cases, $T_i/T_e=2.90$ has either comparable or lower transport than other $T_i/T_e$ values. Further, we find that in both cases, the transport quantities have similar values.}

\end{figure}

\begin{figure}
\includegraphics[width=0.48\textwidth]{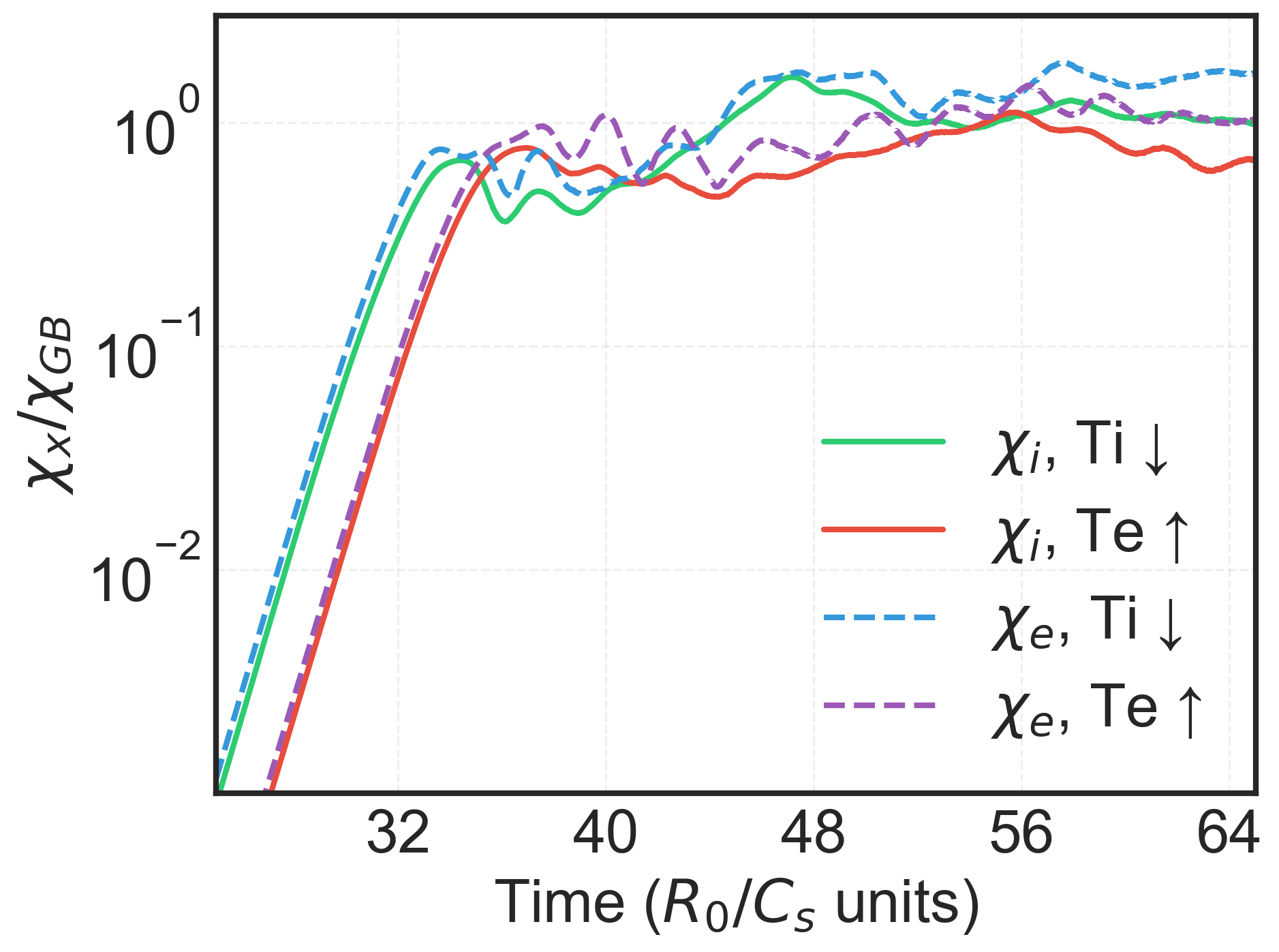}
\includegraphics[width=0.48\textwidth]{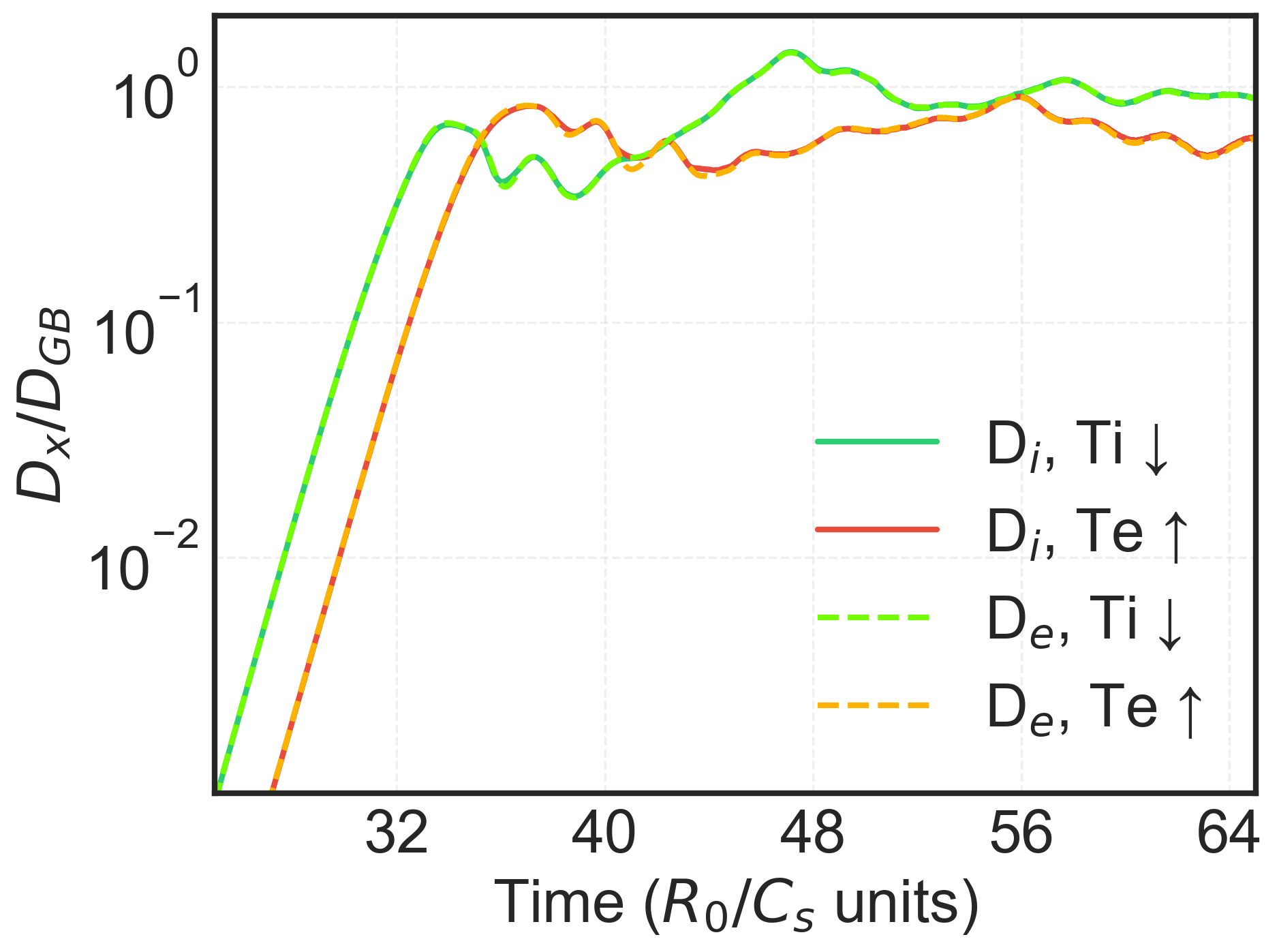}
\centering
\caption{\label{fig:zfcmp1} Comparison of the $\chi_x/\chi_{GB}$ and $D_x/D_{GB}$ {x= i,e} in the presence of ZFs for the case $T_i/T_e=1$ obtained in two ways: decreasing ion temperature at fixed electron temperature and increasing electron temperature at fixed ion temperature. We find that in the case where we increase the electron temperature, the transport quantities have similar values to those in the case where we decrease the ion temperature. We see a similar trend in the electron channel.}

\end{figure}

This large discrepancy in transport enhancement can be attributed to the intrinsic gyro-Bohm scaling of turbulent transport\cite{horton1999drift}. According to this model, diffusivity is expected to scale with the ion gyroradius, $\rho_s \propto \sqrt{T_e}$. This hypothesis is supported by the behavior of the turbulent eddy size, measured via the radial correlation length $L_r$ (Fig.~\ref{fig:correlation}). Increasing $T_e$ results in substantially larger eddies, and hence higher transport, whereas decreasing $T_i$ minimally impacts their size and hence a modest increase in transport in this case\cite{diamond2005zonal,mckee2001non}. To definitively verify the scaling effect, we normalize the transport coefficients by their respective gyro-Bohm values, $\chi_{GB}$ and $D_{GB}$, as shown in Fig.~\ref{fig:tabcmpfig}. We present one such case of $T_i/T_e=1.0$ in Fig~\ref{fig:zfcmp1}. We see that the difference between the electron-heating and ion-cooling cases is largely removed in these dimensionless units. Both scenarios now exhibit comparable saturation levels, while the lowest transport quantities are observed in the original case, where $T_i/T_e=2.90$. 

Having established that the apparent transport differences are a manifestation of gyro-Bohm scaling and that a high $T_i/T_e$ ratio is consistently favorable for confinement, we now turn our investigation to how the inclusion of impurity species—specifically carbon ions—affects microturbulence and transport in the original case of $T_i/T_e=2.90$.

%%%%%%%%%%%%%% Physical Units

\begin{table}

\begin{tabular*}{\textwidth}{@{}l*{4}{@{\extracolsep{0pt plus 12pt}}r}}
\hline
$T_i$/$T_e$ & $2.90$ (Original case) & $1.45$ & $1.00$ & $1/2.90$ \\
\hline
$\chi_i$ (m$^2$/sec) & $4.42$ & $10.58$ & $12.12$ & $10.14$ \\
$\chi_e$ (m$^2$/sec) & $7.90$ & $14.01$ & $14.29$ & $9.53$ \\
$D_i$ (m$^2$/sec) & $3.70$ & $9.07$ & $10.04$ & $7.12$ \\
$D_e$ (m$^2$/sec) & $3.73$ & $9.10$ & $10.06$ & $7.14$ \\
\hline
\end{tabular*}
\caption{\label{tab:satvaliond} Comparison of the saturation values of the transport quantities in the presence of zonal flow for different $T_i/T_e$ ratios for the case when we decrease the ion temperature. In this case, the saturation values do not change significantly compared to the original case of transport quantities.}
\end{table}

%%%%%%%%%%%%%%%%%% GB Units
% \begin{table}
% \caption{\label{tab:satvaliond} Comparison of the saturation values of the transport quantities in the presence of zonal flow for different $T_i/T_e$ ratios for the case when we decrease the ion temperature. In this case, the saturation values do not change significantly compared to the original case of transport quantities.}
% \begin{tabular*}{\textwidth}{@{}l*{4}{@{\extracolsep{0pt plus 12pt}}r}}
% \hline
% $T_i$/$T_e$ & $2.90$ (Original case) & $1.45$ & $1.00$ & $1/2.90$ \\
% \hline
% $\chi_i$  & $0.47$ & $1.04$ & $1.54$ & $1.04$ \\
% $\chi_e$  & $0.74$ & $1.59$ & $2.00$ & $0.97$ \\
% $D_i$  & $0.39$ & $0.92$ & $1.28$ & $0.73$ \\
% $D_e$  & $0.39$ & $0.91$ & $1.28$ & $0.74$ \\
% \hline
% \end{tabular*}
% \end{table}

%%%%%%%%%%%%%%Physical Units

\begin{table}

\begin{tabular*}{\textwidth}{@{}l*{4}{@{\extracolsep{0pt plus 12pt}}r}}
\hline
$T_i$/$T_e$ & $2.90$ (Original case) & $1.45$ & $1.00$ & $1/2.90$ \\
\hline
$\chi_i$ (m$^2$/sec) & $4.42$ & $20.33$ & $33.10$ & $124.06$ \\
$\chi_e$ (m$^2$/sec) & $7.90$ & $26.94$ & $38.15$ & $127.25$ \\
$D_i$ (m$^2$/sec) & $3.70$ & $16.90$ & $29.49$ & $90.18$ \\
$D_e$ (m$^2$/sec) & $3.73$ & $17.05$ & $29.77$ & $89.85$ \\
\hline
\end{tabular*}
\caption{\label{tab:satvalele} Comparison of the saturation values of the transport quantities in the presence of zonal flow for different $T_i/T_e$ ratios for the case when we increase the electron temperature. We find that the saturation values, in this case, change significantly compared to the original case of transport quantities going from $\chi_i=4.42$ m$^2$/sec for $T_i/T_e=2.90$ to $\chi_i=124.06$ m$^2$/sec for $T_i/T_e=1/2.90$.}
\end{table}

%%%%%%%%%%%%%%%%%%%%% Gyro-Bohm Units

% \begin{table}
% \caption{\label{tab:satvalele} Comparison of the saturation values of the transport quantities in the presence of zonal flow for different $T_i/T_e$ ratios for the case when we increase the electron temperature. We find that the saturation values, in this case, do not change significantly compared to the original case of transport quantities going from $\chi_i/\chi_{GB}=0.47$  for $T_i/T_e=2.90$ to $\chi_i/\chi_{GB}=0.49$ for $T_i/T_e=1/2.90$.}
% \begin{tabular*}{\textwidth}{@{}l*{4}{@{\extracolsep{0pt plus 12pt}}r}}
% \hline
% $T_i$/$T_e$ & $2.90$ (Original case) & $1.45$ & $1.00$ & $1/2.90$ \\
% \hline
% $\chi_i$  & $0.47$ & $0.68$ & $0.55$ & $0.64$ \\
% $\chi_e$  & $0.71$ & $1.04$ & $0.69$ & $0.45$ \\
% $D_i$  & $0.48$ & $0.57$ & $0.52$ & $0.44$ \\
% $D_e$ & $0.48$ & $0.56$ & $0.51$ & $0.44$ \\
% \hline
% \end{tabular*}
% \end{table}

\section{\label{sec:impstudy} Addition of impurity to the original profile}
\noindent In this section, we study the effect of adding an impurity species to the original case of $T_i/T_e=2.90$. The carbon atoms are present as impurity ions. In our case, the impurity concentration is $n_z/n_e=0.07$, which is obtained from fitting the experimental data. The ion and the electron temperature are chosen to be the same as the original profile, as in Fig~\ref{fig:tempprofile}. We take the ion temperature,$T_i$, to be the same as the impurity temperature, $T_z$. The reason for $T_z=T_i$ can be seen by comparing the heat transfer time, $\tau_{\alpha \beta}$ from particle species $\beta$ to the particle species $\alpha$ which can be written as\cite{huba20132013}

\begin{eqnarray}
    \tau_{\alpha \beta} \sim \frac{m_{\alpha} m_{\beta}}{Z_{\alpha}^2 Z_{\beta}^2 n_{\beta}} \left( \frac{T_{\beta}}{m_{\beta}} +\frac{T_{\alpha}}{m_{\alpha}} \right)
\end{eqnarray}

Where $m$, $Z$, $T$, and $n$ label the particle mass, charge, temperature, and density, respectively.

The mass of ion, electron, and impurity follows $m_e \ll m_i < m_z$, while all three species' temperatures have the same order of magnitude. Using this approximation, we can simplify the above formula to get the ratio of heat transfer time from electrons to primary ions and main ions to impurity ions as \cite{tao2023numerical}

\begin{eqnarray}
    \frac{\tau_{ie}}{\tau_{zi}} \approx Z_z^2 \frac{n_i}{n_e} \left(\frac{m_i}{m_e}\right)^{1/2} \left(\frac{T_e}{T_i}\right)^{3/2} \frac{m_i}{m_z}
\end{eqnarray}

In the above formula, $Z_z$ is the impurity charge. Now, in our case, $T_i/T_e\sim 2.90$ in the core region of the plasma, $n_i\approx n_e$, $Z_z=6$, and $m_i/m_z=1/12$. Using this, we find that $\tau_{ie}\sim  30\tau_{zi}$. Hence, the time for main and impurity ions to thermalize is less than that for main ions and electrons. This justifies $T_i=T_z$ in the core region of the plasma.

The ion density is modified in the presence of impurities. GTC calculates the ion density by the quasineutrality condition
\begin{eqnarray}
    Z_i n_i= n_e-Z_z n_z
    \label{eq:quasineutrality}
\end{eqnarray}
 Here, $Z_i$ and $n_i$ are the ion charge and mass, $n_e$ is the electron density, and $Z_z$ and $n_z$ are the impurity charge and density, respectively. We choose the electron density in both the presence and absence of impurity as shown in Fig~\ref{fig:denprofile}. The ion density differs depending on whether the impurity is present (represented by an orange dashed line in Fig~\ref{fig:denprofile}) or absent (represented by a green solid line in Fig~\ref{fig:denprofile}).

\begin{table}

\begin{tabular*}{\textwidth}{@{}l*{2}{@{\extracolsep{0pt plus 12pt}}r}}
\hline
 & Without Impurity (Original case) & With Impurity\\
\hline
$\gamma(R_0/C_s$ units) & $0.30$ & $0.41$ \\
$m(k_\theta [cm^{-1}])$ & $35$ (2.4) & $69$ (4.8) \\
$n$ & $17$ & $35$ \\
Mode & ITG & ITG \\
Radial Profile Peak (Location in r/a units) & $0.51$ & $0.51$ \\
Width of Radial Profile (r/a) & $0.06$ & $0.06$ \\
\hline
\end{tabular*}
\caption{\label{tab:linwimp} Comparison of linear simulation of the original case ($T_i/T_e=2.90$) without impurity and with impurity. }
\end{table}

We present the result of the linear regime for the case $T_i/T_e=2.90$ with impurity and without impurity in Table~\ref{tab:linwimp}. The growth increases by adding impurity in the original profile, which suggests that the mode could be an impurity ITG mode\cite{dominguez1993impurity,fulop2006impurity,dong1995studies,li2019impurity,angioni2021impurity}. The radial profile peaks at the exact location in both cases.

\begin{figure}
    \centering
    \begin{overpic}[width=0.48\textwidth]{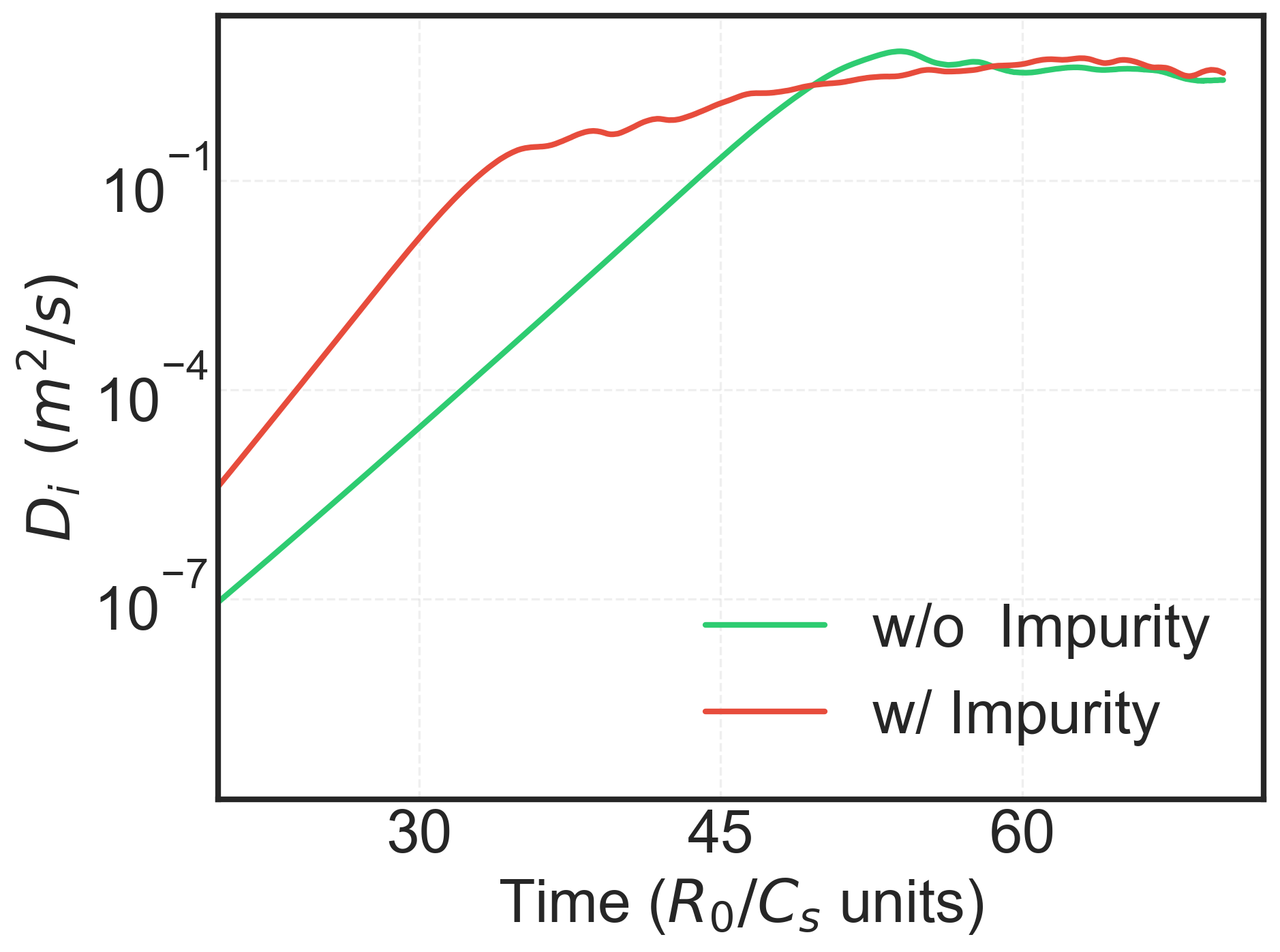}
        % \put(70,50)
    \end{overpic}
    \begin{overpic}[width=0.48\textwidth]{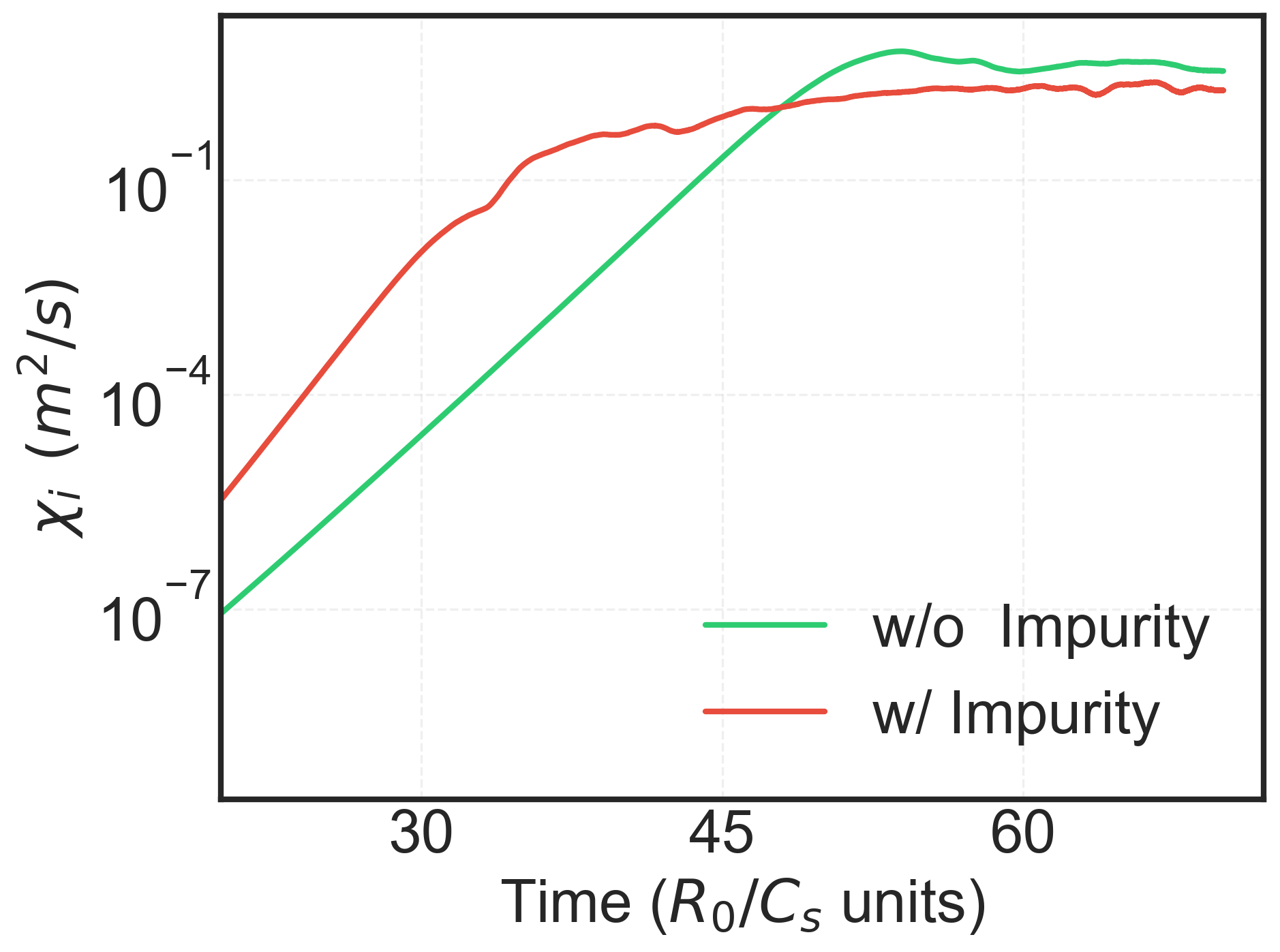}
        % \put(70,50)
    \end{overpic}

    \caption{\label{fig:zfoimp} Effect of ZFs on the particle (\textit{Left}) 
 and heat (\textit{Right}) diffusivities in the presence and absence of impurity for the case $T_i/T_e=2.90$. We see that there is no significant reduction in the ion transport quantities. The trend is similar for the case of electron heat and particle transport.}
\end{figure}

\begin{figure}
\includegraphics[scale=0.45]{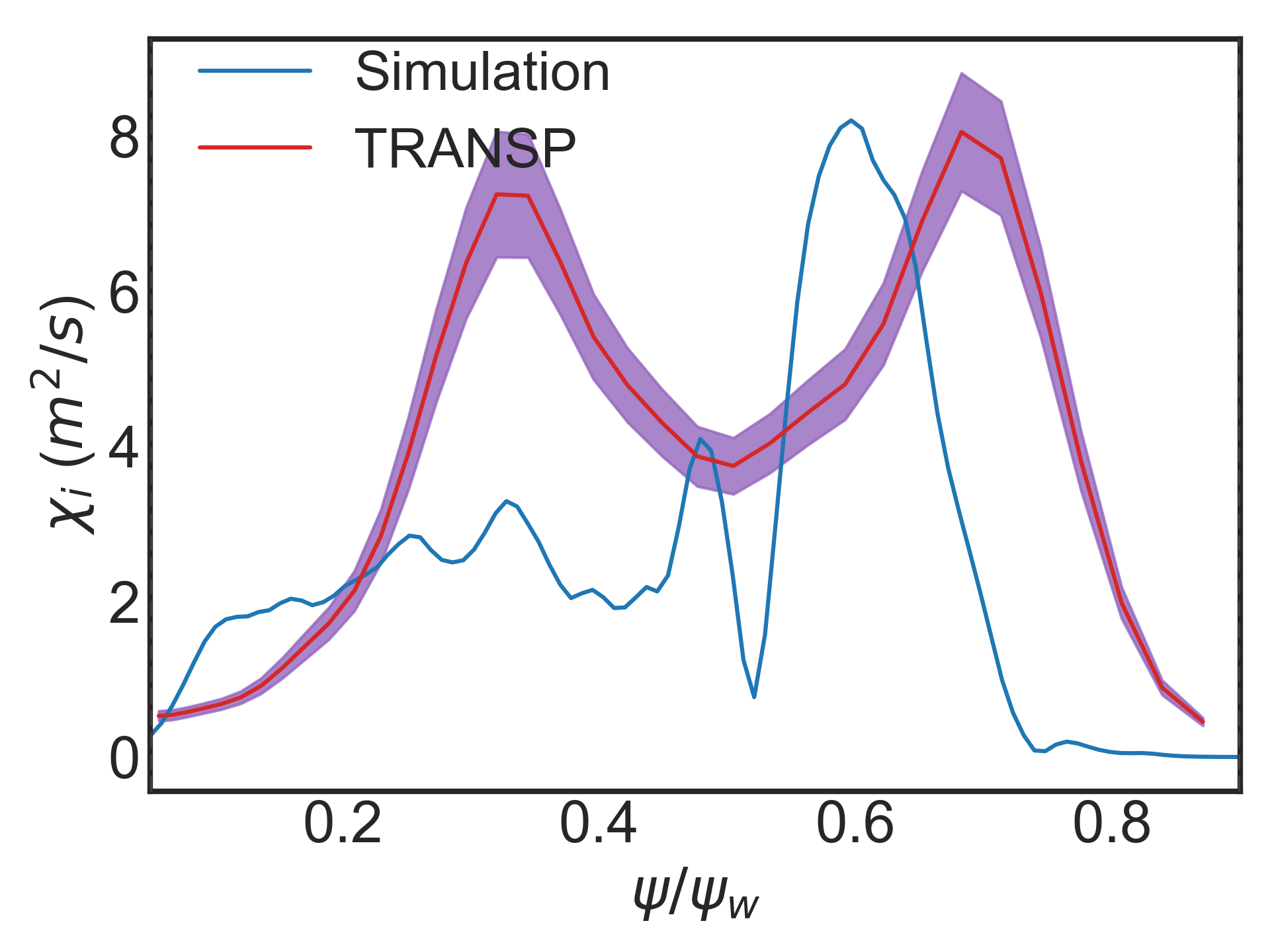}
\includegraphics[scale=0.45]{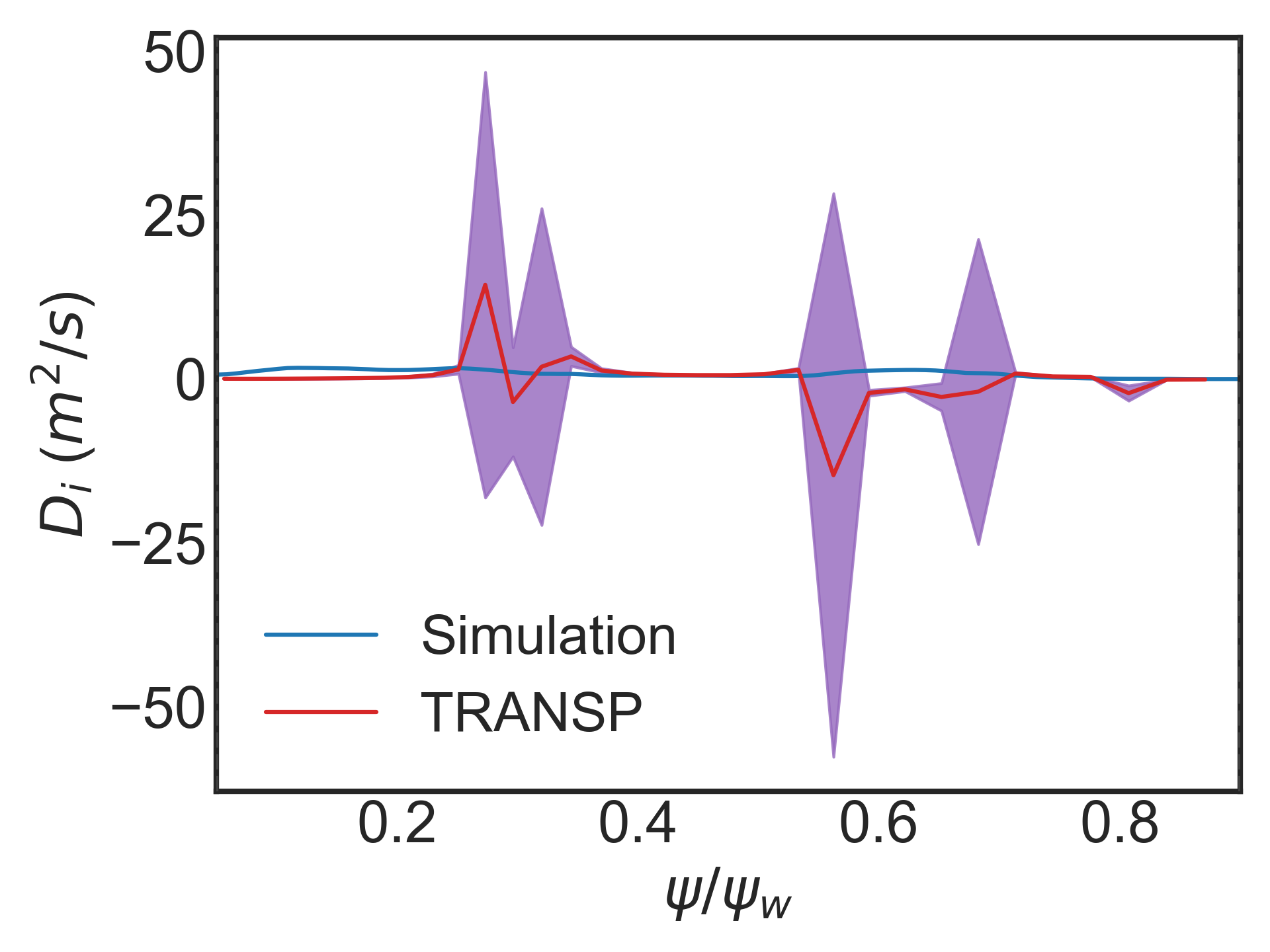}
\centering
\caption{\label{fig:transpion} Comparison of the transport in the ion channel for the original case temperature ratio $T_i/T_e=2.90$ with impurity simulation result with the TRANSP code. Our simulation results are within the ballpark limit of the result predicted by the TRANSP code.}

\end{figure}

\begin{figure}
\includegraphics[scale=0.45]{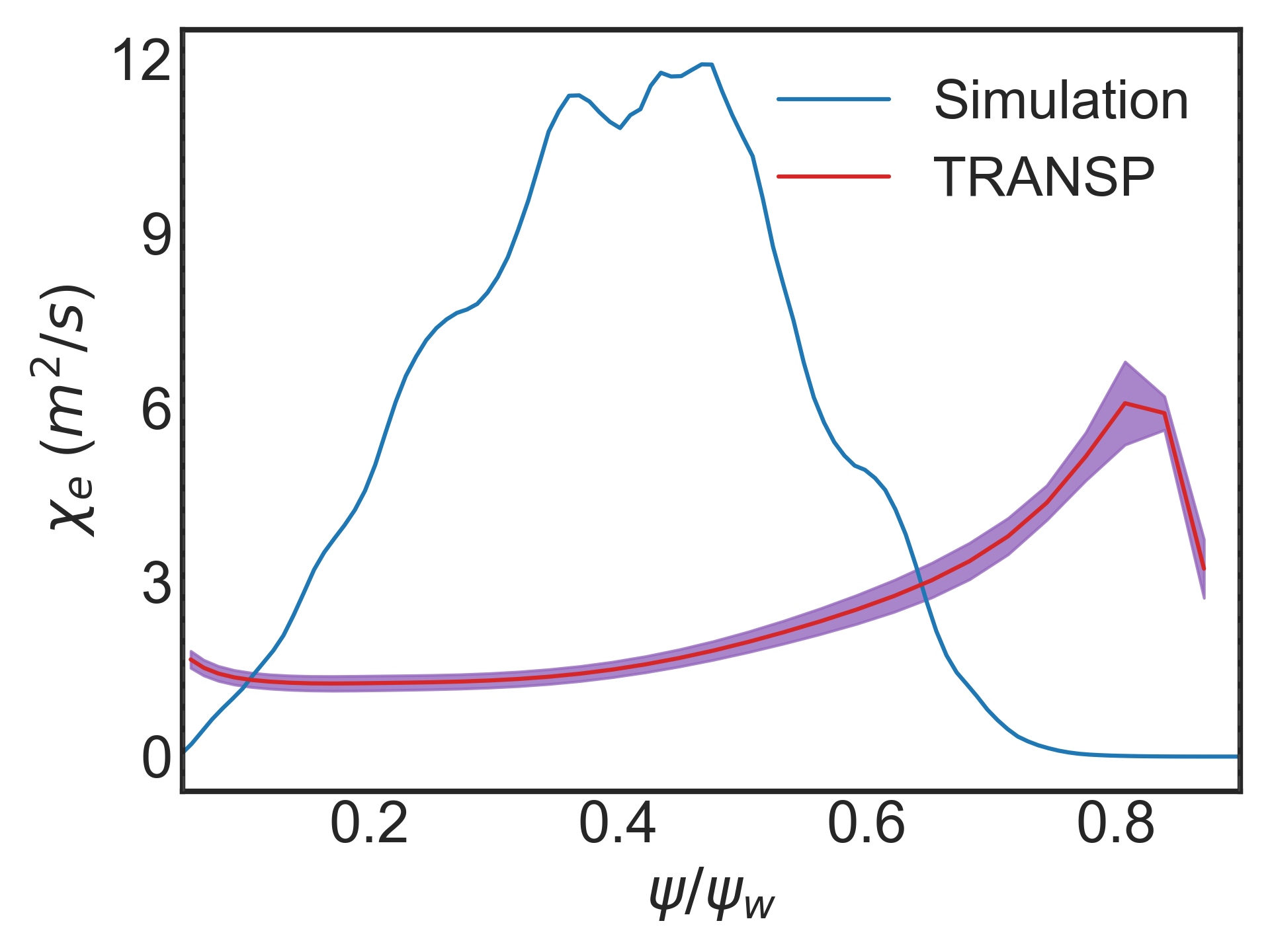}
\includegraphics[scale=0.45]{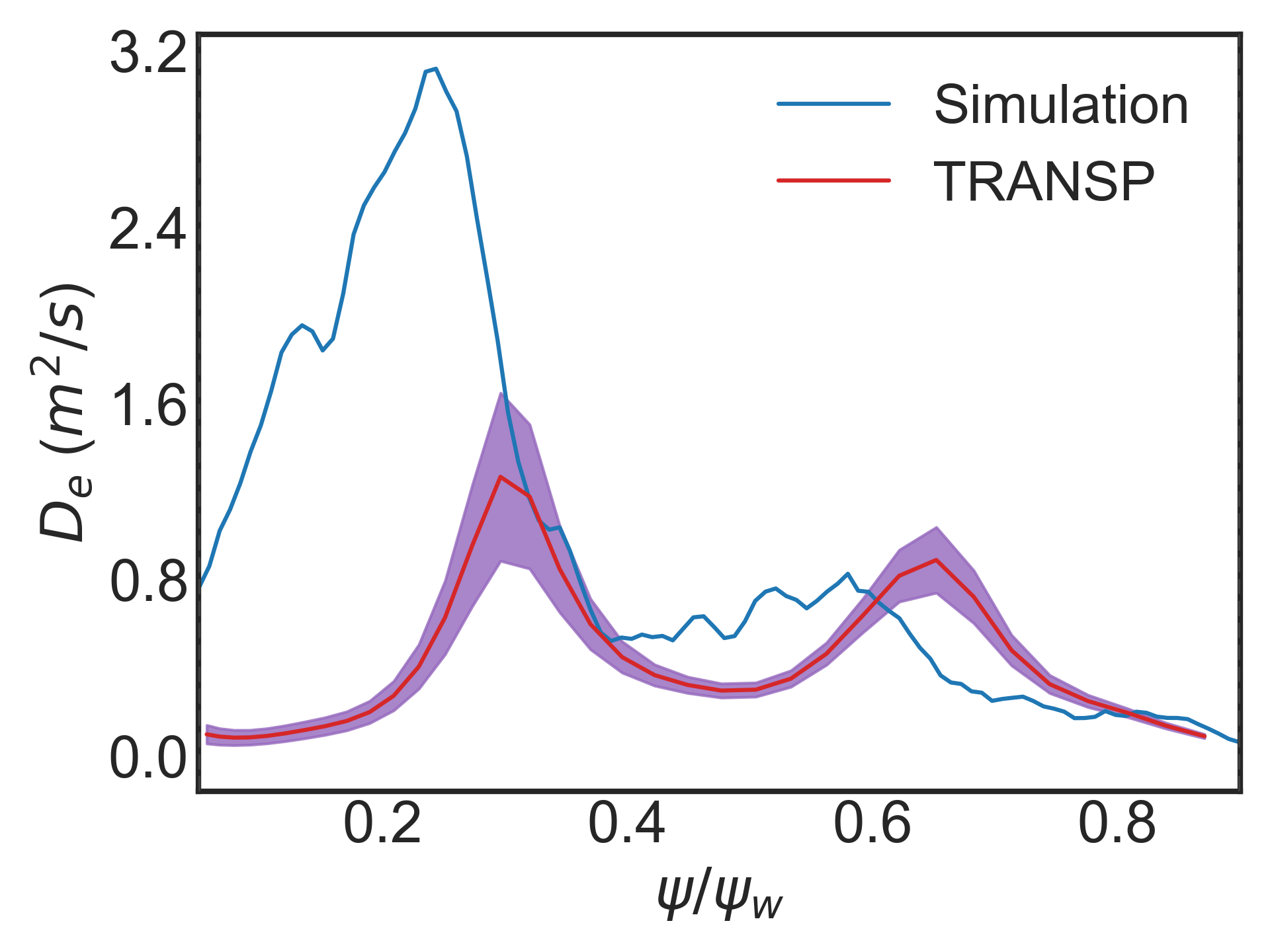}
\centering
\caption{\label{fig:transpelectron} Comparison of the transport in the electron channel for the original case temperature ratio $T_i/T_e=2.90$ with impurity simulation result with the TRANSP code. While the two have quantitative differences, the overall trends exhibit similarities, providing valuable insights into the underlying transport mechanisms.}

\end{figure}

We now present the result of the nonlinear regime for this case. We compare the effect of ZFs for the original profile with and without impurity. We see that in Fig.~\ref{fig:zfoimp} (a) and (b), there is no significant reduction in the transport quantities. However, the saturation level is reached earlier in the presence of impurity than without impurity (In the presence of impurity, the saturation level is achieved at $t=35 R_0/C_s$ while in the absence of impurity, it is reached at $t=53 R_0/C_s$). These results can be understood from the behavior of the zonal flow amplitude, which initiates earlier in the presence of impurity. This earlier onset of zonal flow activity explains why the nonlinear saturation is reached sooner compared to the case without impurity.

 \noindent To further validate these impurity-driven simulation results, we compare the transport levels with those obtained from the TRANSP code \cite{osti_1489900}. We see that the transport quantities in the ion channel, as computed by the GTC code, are lower than the TRANSP results. Nonetheless, they are in the same order of magnitude. Moreover, the strong gradients in the pedestal region may influence transport in the core region, an effect that is not captured in the present simulations\cite{ida2022non,hahm2005dynamics}. The remaining differences may also arise from the effects of radial electric fields and finite beta, which are not included in the current simulations. In addition, the radial profiles of the electron heat diffusivity exhibit qualitative differences: the simulation predicts higher transport in the mid-radius region, while the TRANSP results show enhanced transport toward the edge. Similar trends have been reported in earlier L-mode gyrokinetic validation studies on DIII-D \cite{holland2011advances}, where local $\delta f$ simulations underpredicted transport near the edge and overpredicted it in the mid-radius region. These discrepancies were attributed to the neglect of self-consistent $E\times B$ shear, the strong stiffness of the temperature profiles, and the absence of nonlocal (core–edge coupled) dynamics. In our case, the exclusion of a self-consistent pedestal and these stabilizing effects could lead to an underestimation of edge transport and a relative overprediction at mid-radius. Future electromagnetic and core–edge integrated simulations are expected to address this discrepancy more comprehensively.

We now extend our study to explore how different ion compositions influence QH-mode plasmas. In particular, since ITER plans to operate with helium during its initial non-nuclear phase\cite{ITER2024}, it is essential to understand the impact of using helium instead of hydrogen on turbulence and transport dynamics.

\begin{table}

\begin{threeparttable}
\begin{tabular*}{\textwidth}{@{}l*{4}{@{\extracolsep{0pt plus 12pt}}l}}
\hline
$T_i$ & $10.24$ keV & $7.06$ keV & $3.54$ keV & $1.22$ keV \\
\hline
$T_i/T_e$ & $2.90$ (Original case) & $1.45$ & $1$ & $1/2.90$ \\
\hline
m $(k_\theta [cm^{-1}])$ & $71 (5.0)$ & $96 (6.7)$ & $104 (7.5)$ & $128 (8.9)$ \\
n & $36$ & $48$ & $53$ & $65$ \\
$\gamma (R_0/C_s$ units) & $0.63$ & $0.79$ & $0.80$ & $0.79$ \\
Mode & ITG & TEM & TEM & TEM \\
% Width (r/a units) & $0.059$ & $0.082$ & $0.096$ & $0.126$  \\
Width (r/a units) & $0.06$ & $0.08$ & $0.10$ & $0.12$  \\
\hline
\end{tabular*}
\caption{\label{tab:helinion1} Comparison of the various quantities in the linear simulation when we decrease the ion temperature, keeping the electron temperature fixed at $3.54$ keV when helium is present as the main ion species.}
 % \begin{tablenotes}
 %    \item[*] Inconclusive Case.
 %  \end{tablenotes}
\end{threeparttable}
\end{table}

\begin{table}

\begin{threeparttable}
\begin{tabular*}{\textwidth}{@{}l*{4}{@{\extracolsep{0pt plus 12pt}}l}}
\hline
$T_e$ & $3.53$ keV & $5.13$ keV & $10.24$ keV & $29.59$ keV \\
\hline
$T_i$/$T_e$ & $2.90$ (Original case) & $1.45$ & $1$ & $1/2.90$ \\
\hline
m $(k_\theta [cm^{-1}])$ & $71 (5.0)$ & $73 (5.1)$ & $57 (4.0)$ & $43 (3.0)$ \\
n & $36$ & $37$ & $29$ & $21$ \\
$\gamma (R_0/C_s$ units) & $0.67$ & $0.71$ & $0.70$ & $0.70$ \\
Mode & ITG & ITG & TEM & TEM \\
% Width (r/a units) & $0.059$ & $0.082$ & $0.067$ & $0.104$  \\
Width (r/a units) & $0.06$ & $0.08$ & $0.07$ & $0.10$  \\
\hline
\end{tabular*}
\caption{\label{tab:helinele1} Comparison of the various quantities in the linear simulation for the case when the electron temperature increases, keeping the ion temperature fixed at $10.24$ keV when the helium is present as the main ion species.}
 % \begin{tablenotes}
 %    \item[*] Inconclusive Case.
 %  \end{tablenotes}
\end{threeparttable}
\end{table}

\section{\label{sec:heplasma} Turbulence and transport in core QH-mode plasmas with Helium as main ion species}

Recently, the ITER Research Plan\cite{ITER2024} has proposed a new strategy to use either hydrogen(H) or helium(He) or both of them in a staged manner during the first operational step. Helium is produced in fusion reactors through the deuterium-Tritium reaction $^2$H+$^3$H $\rightarrow ^4$He+n and the concentration can reach a few percent\cite{hakola2024helium}. Additionally, He can be used in the initial, non-nuclear operational phase of a reactor to keep nuclear activation at low levels. ITER aims to access helium H-mode in the initial stages due to the lower L-H power threshold observed in Helium plasmas, which can help ITER operation starting at lower external power levels\cite{ITER2024}. However, the presence of ELMs and challenges with the edge-core integration issues may hinder the sustainment of H-mode plasma in helium. As discussed earlier, QH-mode plasma offers a promising alternative to H-mode, as it is naturally ELM-free. However, QH-mode plasmas in helium have not yet been explored in the literature. We provide a short discussion on the impact of helium as the main ion species on core transport in this paper.

% As discussed in Sec~\ref{sec:intro}, ITER is planning to use helium in the initial, non-nuclear operation phase of the reactor\cite{ITER2024}. 

Motivated by these considerations, we aim to study how QH-mode plasma behaves in the presence of helium when ICRH or ECH is applied. To simulate the effect of these heating mechanisms, we analyze the $T_i/T_e$ ratio as in the previous section. We keep all the parameters($T_i$, $T_e$, and $n_e$) be same as presented in Fig~\ref{fig:tempprofile} and Fig~\ref{fig:denprofile}. However, due to the quasineutrality condition(Eq. [~\ref{eq:quasineutrality}]), $n_i$ is equal to half the value of $n_e$.

\begin{table}

\begin{tabular*}{\textwidth}{@{}l*{4}{@{\extracolsep{0pt plus 12pt}}r}}
\hline
$T_i$/$T_e$ & $2.90$ (Original case) & $1.45$ & $1.00$ & $1/2.90$ \\
\hline
$\chi_i$ (m$^2$/sec) & $7.73$ & $7.28$ & $8.12$ & $4.43$ \\
$\chi_e$ (m$^2$/sec) & $7.09$ & $7.35$ & $8.61$ & $4.46$ \\
$D_i$ (m$^2$/sec) & $6.94$ & $5.94$ & $6.41$ & $3.04$ \\
$D_e$ (m$^2$/sec) & $6.97$ & $5.96$ & $6.43$ & $3.05$ \\
\hline
\end{tabular*}
\caption{\label{tab:hesatvaliond} Comparison of the saturation values of the transport quantities in the presence of zonal flow for different $T_i/T_e$ ratios for the case when we decrease the ion temperature and helium is present as the main ion species. In this case, the saturation values do not change significantly compared to the original case of transport quantities.}
\end{table}

\begin{table}

\begin{tabular*}{\textwidth}{@{}l*{4}{@{\extracolsep{0pt plus 12pt}}r}}
\hline
$T_i$/$T_e$ & $2.90$ & $1.45$ & $1.00$ & $1/2.90$ \\
\hline
$\chi_i$ (m$^2$/sec) & $7.73$ & $11.12$ & $23.65$ & $89.44$ \\
$\chi_e$ (m$^2$/sec) & $7.09$ & $10.68$ & $21.78$ & $74.25$ \\
$D_i$ (m$^2$/sec) & $6.94$ & $9.33$ & $18.40$ & $57.63$ \\
$D_e$ (m$^2$/sec) & $6.97$ & $9.41$ & $18.49$ & $57.52$ \\
\hline
\end{tabular*}
\caption{\label{tab:hesatvalele} Comparison of the saturation values of the transport quantities in the presence of zonal flow for different $T_i/T_e$ ratios for the case when we increase the electron temperature in the presence of helium as the main ion species. We find that the saturation values, in this case, change significantly compared to the original case of transport quantities going from $\chi_i=4.42$ m$^2$/sec for $T_i/T_e=2.90$ to $\chi_i=124.06$ m$^2$/sec for $T_i/T_e=1/2.90$.}
\end{table}

We first study the effect of decreasing $T_i$ at the fixed $T_e$. Table~\ref{tab:helinion1} shows the linear regime of the nonlinear simulation. We see that the poloidal and toroidal mode numbers increase as $T_i/T_e$ ratio is increased, same trend as for deuterium case. Since $m\sim k_y\rho_s$ and $\rho_s\sim\sqrt{m_i}$\cite{horton1999drift}, we expect the poloidal and toroidal mode numbers to be higher than those in Table~\ref{tab:linion1}. Furthermore, we observe a higher growth rate in the helium plasma. Additionally, the growth rate saturates as 
 $T_i/T_e$ is decreased from $1.45$ to $1/2.90$. The frequency also changes from ITG to TEM when deuterium is used as the main ion species. Next, we examine the nonlinear regime for this scenario. As shown in Table~\ref{tab:hesatvaliond}, we see that the saturation values of the transport quantities do not change significantly from the original case. Also, the transport values are similar to those in the Table~\ref{tab:satvaliond}.

We next increase $T_e$ to modify the $T_i/T_e$ ratio. Table~\ref{tab:helinele1} shows the linear regime of the full nonlinear simulation. We see that the poloidal and toroidal mode numbers decrease overall and are higher than the corresponding mode numbers in Table~\ref{tab:linele1}. The growth rate first increases and then saturates. Furthermore, the growth rates are higher than those observed in the corresponding deuterium case. The ITG mode is relatively stabilized, and the TEM mode is destabilized when the $T_i/T_e$ ratio is decreased, as when deuterium was used as the main ion species. Next, we examine the nonlinear regime for this scenario. Table~\ref{tab:hesatvalele} shows that the saturation values increase as $T_i/T_e$ is increased, as in the case for deuterium in Table~\ref{tab:satvalele}. Further, the saturation quantities have slightly lower values than when deuterium is used as the main ion species.

Overall, our findings highlight that while helium QH-mode plasmas exhibit differences in 
microinstability behavior and transport characteristics compared to deuterium, their fundamental 
stability properties remain largely intact. The lower saturation transport values in helium 
plasmas suggest potential advantages in confinement, particularly in early-stage ITER operations. However, the practical use of helium in devices with tungsten plasma-facing components has raised concerns about materials and plasma–material interaction (PMI), particularly regarding helium exposure, which can promote the formation of nanoscale “fuzz” on tungsten surfaces. These issues are actively being evaluated in the ITER planning and PMI communities. Consequently, while helium remains a useful test case for studying Ti/Te effects and QH-mode physics, extrapolation to ITER-scale operation should be made with caution and in the context of ongoing PMI considerations \cite{kajita2020tungsten, ITER2024}.

\section{\label{sec: conclude} Conclusion and future work}

In this study, we have investigated the impact of the ion-to-electron temperature ratio ($T_i/T_e$) on microturbulence-driven transport in Quiescent H-mode (QH-mode) plasmas using gyrokinetic simulations with the GTC code. Our analysis reveals that a reduction in $T_i/T_e$
leads to a relative destabilization of TEM over the ITG modes, accompanied by increased linear growth rates. We also find elevated transport saturation levels under conditions of increasing $T_e$ at fixed $T_i$. Conversely, decreasing $T_i$ at fixed $T_e$ also leads to an ITG-to-TEM transition but results in more modest increases in transport levels, typically up to $3\times$. Zonal flows play a crucial role in suppressing transport, with the most potent suppression occurring at high $T_i/T_e$. The correlation length analysis further reveals that radial eddy sizes increase with rising $T_e$ and decrease slightly with falling $T_i$, consistent with the trends observed in transport saturation in physical units. Additionally, our analysis of impurity effects reveals that impurity presence does not substantially alter transport quantities, although it influences nonlinear saturation dynamics.

Furthermore, we examined the impact of using helium as the main ion species, motivated by ITER’s initial operational plans. Our findings indicate that helium QH-mode plasmas exhibit similar turbulence trends to deuterium plasmas but with distinct differences in mode growth rates and transport saturation levels. While helium plasmas experience higher growth rates, they also demonstrate lower saturation transport values, which may offer advantages for confinement. These insights contribute to the broader understanding of turbulence suppression mechanisms and confinement optimization in QH-mode plasmas.

Overall, this study provides valuable guidance for future experimental and theoretical research to optimize plasma performance in next-generation fusion devices where different heating mixtures (predominantly electron-heated plasmas like ITER and SPARC) will be used, leading to operations at lower $T_i/T_e$. Further work incorporating additional heating schemes, collisional effects, and experimental validation will be essential to refine our understanding of QH-mode stability and transport in deuterium and helium plasmas.

Future work will study the effects of ExB shear and finite-$\beta$ on mode growth rates as well as transport fluxes driven by the unstable modes. Sensitivity scans of ion and electron temperature gradients will be performed to understand the proximity to critical gradients, which can drive higher nonlinear fluxes. QH-mode plasmas of DIII-D with additional ECH power at constant NBI power are currently being investigated, where a $T_i/T_e$ value of 1.2-2.7 is obtained. The measured density turbulence character and its amplitude evolution with varying $T_i/T_e$ will be compared to these simulations. The present work will be extended to understand the turbulence and transport in the edge region of the QH-mode plasmas. Future work will also investigate the interaction of large-scale EHOs with that of small-scale microturbulence.    

\section*{Acknowledgments}
This material is based upon work supported by the U.S. Department of Energy, Office of
Science, Office of Fusion Energy Sciences, using the DIII-D National Fusion Facility, a DOE
Office of Science user facility, under Award(s) DE-SC0022563, DE-SC0019352, and DE-FC02-04ER54698. This work is supported by the Board of Research in Nuclear Sciences (BRNS Sanctioned no. and 57/14/04/2022-BRNS), Science and Engineering Research Board EMEQ program (SERB sanctioned no. EEQ/2022/000144), National Supercomputing Mission (NSM), US Department of Energy under Award No.DE-SC0024548 and DE-FG02-07ER54916. We acknowledge National Supercomputing Mission (NSM) for providing computing resources of `PARAM PRAVEGA' at S.E.R.C. Building, IISc Main Campus Bangalore, which is implemented by C-DAC and supported by the Ministry of Electronics and Information Technology (MeitY) and Department of Science and Technology (DST), Government of India and  ANTYA cluster at Institute of Plasma Research, Gujarat, and by US DOE SciDAC and INCITE. A.T. thanks the University Grants Commission (UGC) for supporting him as a Senior Research Fellow (SRF).\\

\noindent{Disclaimer}: This report was prepared as an account of work sponsored by an agency of the
United States Government. Neither the United States Government nor any agency thereof, nor
any of their employees, makes any warranty, express or implied, or assumes any legal liability or responsibility for the accuracy, completeness, or usefulness of any information, apparatus,product, or process disclosed, or represents that its use would not infringe privately owned rights. Reference herein to any specific commercial product, process, or service by trade name,trademark, manufacturer, or otherwise does not necessarily constitute or imply its endorsement,recommendation, or favoring by the United States Government or any agency thereof. The views and opinions of authors expressed herein do not necessarily state or reflect those of the United States Government or any agency thereof.

\bibliography{Untracked_bib2}

\vspace{0.5cm}
\begin{center}
\line(1,0){200}
\end{center}

%\textbf{Keywords:} Finite element method; Poisson equation; Global toroidal code
\end{document}